\documentstyle[12pt,aaspp4]{article}
\begin{document}
\parindent=1.0cm

\title{Deep ALTAIR$+$NIRI Imaging of the Disk and Bulge of M31
\footnote[1]{Based on observations obtained at the
Gemini Observatory, which is operated by the Association of Universities
for Research in Astronomy, Inc., under a co-operative agreement with the
NSF on behalf of the Gemini partnership: the National Science Foundation
(United States), the Particle Physics and Astronomy Research Council
(United Kingdom), the National Research Council of Canada (Canada),
CONICYT (Chile), the Australian Research Council (Australia), CNPq (Brazil),
and CONICET (Argentina).}}

\author{T. J. Davidge}

\affil{Canadian Gemini Office, Herzberg Institute of Astrophysics,
\\National Research Council of Canada, 5071 West Saanich Road,
\\Victoria, B.C. Canada V9E 2E7\\ {\it email: tim.davidge@nrc.ca}}

\author{K. A. G. Olsen \& R. Blum}
\affil{Cerro Tololo Inter-American Observatory, 
\\National Optical Astronomy Observatory, Casilla 603, 
\\La Serena, Chile\\ {\it email: kolsen@noao.edu, rblum@noao.edu}}

\author{A. W. Stephens}
\affil{Princeton University, Peyton Hall, Ivy Lane, Princeton, NJ 08544-1001,
\\{\it email: stephens@astro.princeton.edu}}

\author{F. Rigaut}
\affil{Gemini Observatory, 670 North A'ohoku Place,
\\Hilo, HI 96720-2700\\ {\it email: frigaut@gemini.edu}}

\begin{abstract}

	Deep $J, H,$ and $K'$ images, recorded with the ALTAIR adaptive optics 
system and NIRI imager on Gemini North, are used to probe the stellar content 
of the disk and bulge of the Local Group galaxy M31. With FWHM 
near 0.08 arcsec in $K$, these are the highest angular resolution 
near-infrared images yet obtained of this galaxy. One field samples the outer 
disk of M31 at a galactocentric distance of roughly 62 arcmin along the major 
axis. The mean metallicity in this field is close to that of the metal-rich 
globular cluster NGC 6528, and no stars with [Fe/H] $< -0.7$ are detected. 
Another field, located on the major axis 9 arcmin from the galaxy center, 
contains a roughly equal mix of disk and bulge stars. 
The RGB in this field is redder than that of NGC 6528, although it is argued 
that reddening internal to M31 may be significant in this region of the galaxy. 
The remaining two fields, located at projected galactocentric 
distances of 2 and 4 arcmin, are dominated by bulge stars. 
The RGB-tip occurs between $K = 17.0$ and 17.2, and the color of the RGB in 
the field closest to the center of M31 is consistent with that of NGC 
6528. After accounting for random photometric errors, the upper 
RGB in each field has a width on the $(K, J-K)$ CMDs that 
is consistent with a $\pm 0.5$ dex dispersion 
in [Fe/H], in rough agreement with what is seen in other disk 
and spheroid fields in M31. The number of bright AGB and RGB stars also scales 
with the $r-$band surface brightness in all four fields. Thus, we conclude 
that the stellar content does not change markedly from field-to-field, 
and that the photometric properties of the brightest AGB stars in 
the two innermost fields are not affected significantly 
by crowding. The brightest star has M$_K = -8.6$ and M$_{bol} = -5.2$, although 
this may not be a reliable measure of the AGB-tip brightness due to photometric 
variability.

	A population of very bright red stars, which 
we identify as C stars, are seen in the three fields that are closest to the 
center of M31. The spatial distribution of these objects suggest 
that they are well mixed throughout this part of M31, and so 
likely did not form in a compact region near the galactic nucleus, but 
more probably formed in the inner disk. We speculate 
that these C stars may be the most luminous members of the intermediate 
age population that has been detected previously in studies of the 
integrated spectrum of the central regions of M31.

\end{abstract}

\keywords{galaxies: individual (M31) - galaxies: stellar content - stars: RGB - stars: AGB and post-AGB}

\section{INTRODUCTION}

	In recent years considerable effort has been dedicated to determining 
when the morphological characteristics of galaxies were set in place. In the 
case of spiral galaxies a major issue is the relative timescale of 
bulge and disk formation (e.g. Bouwens, Cayon, \& Silk 1999). 
One class of model argues that bulges form 
early-on, prior to the creation of stable disks, as 
the result of mergers or a monolithic collapse. Indeed, cosmological 
models that assume a cold dark matter (CDM) dominated Universe predict that 
large galaxies formed through the accretion of small gas-rich proto-galactic 
fragments. These fragments were likely short-lived, as they were consumed in 
major mergers that altered morphologies through kinematic heating and 
spurred large-scale episodes of star formation 
(e.g. Somerville, Primack, \& Faber 2001). Given the chaotic state of the early 
Universe, and the difficulty in maintaining stable disks in such 
environments (e.g. Weil, Eke, \& Efstathiou 1998), it 
might be anticipated that bulges were the first long-lived structures to 
form in the systems that would eventually become spiral galaxies. This 
is consistent with the relative colors of bulges and 
disks in the Hubble Deep Field (Abraham et al. 1999), and with the age 
of the Galactic bulge (e.g. Feltzing \& Gilmore 2000, Ortolani et al. 1995).

	At the other extreme are models in which disks formed first, 
and bulges are the result of secular processes involving disk material. 
The motion of gas in the disk can lead to the formation of spheroidal 
structures, and bars, which may form when gas is channeled towards the 
central regions of a galaxy, are key elements in 
this process. The bar transports angular momentum from the 
central regions of the galaxy to larger radii, and stars that formed in the bar 
will be scattered out of the disk to form a spheroidal structure when the bar 
buckles (e.g. Friedli \& Benz 1995). Stars that form in the 
inner regions of a galaxy in the absence of a bar may also be 
scattered into the surrounding bulge by interactions with molecular clouds 
(Kim \& Morris 2001). Consequently, the inner bulges of systems that 
are experiencing, or did experience, nuclear star formation might contain
a diffuse population of intermediate age stars that did not form {\it in situ}.

	Evidence that material originally in the disk contributes to 
the formation of bulges comes from the structural properties of the spheroids 
in late-type galaxies. Rather than following an R$^{1/4}$ profile that is 
the expected signature of violent relaxation, 
the central spheroids of late-type spiral galaxies follow Sersic 
profiles with exponents that are indicative of a more leisurely formation 
process (e.g. Balcells et al. 2003; Courteau, de Jong, \& Broeils 1996; 
Andredakis, Peletier, \& Balcells 1995). It might also be anticipated that the 
stellar contents of bulges will show a morphological dependence if they are 
built up by secular processes, in the sense that the luminosity-weighted ages 
of bulges will become younger as one moves to later morphological types. 
Nevertheless, some exponential bulges have colors that are similar to those 
of bulges that follow R$^{1/4}$ light profiles (Carollo et al. 2001), 
suggesting that secular evolution may have occured 
early-on, but was subsequently curtailed. The development of bulges from disk 
material may also affect the large scale properties of disks, as the mixing 
caused by bars will smooth out pre-existing abundance 
profiles in the disk (e.g. Friedli \& Benz 1995).

	The monolithic collapse and secular evolution models of bulge formation 
make certain predictions about the properties of bulges and their component 
stars. If bulges form early-on in violent collapse episodes then they will be 
made up of old stars. Metallicity gradients that are the 
signatures of dissipation should also be present. Finally, the chemical 
contents of stars formed in such a manner will show the signatures 
of rapid enrichment. On the other hand, bulges that are assembled solely from 
secular processes will be comprised of young or intermediate age stars. 
There may be an age gradient, in the sense that the youngest stars 
are located closest to the disk plane. The chemical 
compositions of stars in a bulge assembled in such a way will 
also show the signatures of a slow enrichment history. 

	As noted by Bouwens et al. (1999), reality likely falls somewhere 
between the extremes defined by the monolithic collapse and secular evolution 
models. Indeed, bulges almost certainly do not evolve in isolation, and there 
are observational indications that the evolution of bulges and disks are 
coupled, even long after the bulge is in place. There is a high incidence of 
distinct nuclear sources in the central regions of nearby bulges, which have 
spectrophotometric properties that differ from those of the surrounding bulge 
and show evidence of recent star formation (e.g. Boker et al. 1999, 2002; 
Carollo et al. 2002). Reservoirs of cool gas must be present near the centers 
of these galaxies, and the surrounding disks are a likely source of this 
material. This interplay between disks and bulges has been on-going for some 
time, as central blue knots are seen in many disk galaxies at 
intermediate redshifts (Ellis, Abraham, \& Dickinson 2001).

	The observational challenges inherent to studies of very distant 
systems can make it difficult to use them to distinguish between the 
various processes that affect bulge evolution (e.g. Bouwens et al. 1999). 
However, the interplay between disks and bulges will have 
observational consequences in nearby spiral galaxies, and so these systems are 
key laboratories for testing models of bulge evolution. 
Studies of the resolved stellar contents in the bulges of nearby galaxies 
can be used to search for the young and intermediate age stars that are 
the expected signatures of secular processes. Moreover, if stars on the RGB 
are resolved then it may be possible to search for the abundance gradients 
that are the expected signatures of a dissipational collapse.

	As the closest external spiral galaxy, M31 is an obvious target for 
detailed study. There are hints that, like many other nearby 
systems, the central regions of M31 may have 
experienced star formation in the not too distant past. The colors of the
nucleus differ from those of the surrounding bulge (King, Stanford, \& Crane 
1995; Davidge et al. 1997; Brown et al. 1998; Lauer et al. 1998), 
suggesting a distinct stellar content, such as might 
remain after the termination of star formation during the last few Gyr. 
There are also spectroscopic signatures of a central young and/or 
intermediate age population (e.g. Bica, Alloin, \& Schmidt 1990; Davidge 
1997; Sil'chenko, Burenkov, \& Vlasyuk 1998), although a population of resolved 
intermediate-aged stars has yet to be detected in the inner regions of M31. 
In fact, the brightest red stars are well-mixed throughout the inner bulge, 
and so likely formed with the main body of the bulge (Davidge 2001). In 
addition, the brightnesses and densities of luminous AGB stars in the bulge of 
M31 are similar to those in systems that are as disparate as M32 and NGC 5128, 
further supporting the notion that they come from an old population 
(Davidge 2002). The red giant branch of the M31 bulge has a color that is 
consistent with a near-solar metallicity (Jablonka et al. 1999; Stephens 
et al. 2001), although the RGB-tip may be depressed with respect to 
larger galactocentric distances (Jablonka et al. 1999). This latter result is 
counter to what would be expected if metallicity decreases with increasing 
galactocentric distance, but could be indicative of an age gradient. 

	Additional hints into the star forming history of the 
central regions of M31 come from studies of stars in the disk of the galaxy. 
Morrison et al. (2004) find a population of globular clusters 
in M31 that have disk kinematics and span a range of metallicities, 
suggesting that a stable disk was in place at very early epochs, so that 
secular interactions between the disk and bulge may have started early-on. 
Young and intermediate age stars are seen at moderate to large 
galactocentric distances (e.g. Richer, Crabtree, \& Pritchet 1990; 
Davidge 1993; Brewer, Richer, \& Crabtree 1995; Cuillandre et al. 
2001; Sarajedini \& Van Duyne 2001; Ferguson \& Johnson 2001), and 
there was large scale star formation throughout much 
of the disk up to 1 Gyr in the past (Williams 2002). However, 
star formation at present appears to be restricted to the spiral arms, where 
the star formation rate has not changed during the past Gyr (Williams 2002). 
The metallicity distribution is remarkably similar over a large range of 
galactocentric distances, peaking near [Fe/H] $= -0.6$, with only a very modest 
number of stars with [Fe/H] $< -1.5$, even in fields that sample the halo 
(e.g. Bellazzini et al. 2003). Finally, there have been interactions between 
M31 and its companions (e.g. Saito \& Iye 2000; Choi, Guhathakurta, \& Johnston 
2002; Ferguson et al. 2002), which have probably influenced the 
star-forming history of M31, and may have triggered flows of gas into the 
central regions of the galaxy. 

	Are there spatially resolved young or intermediate age 
stars in the innermost regions of M31? Do studies of the photometric properties 
of resolved RGB stars reveal signs of a metallicity gradient 
in the inner spheroid, as predicted if the bulge experienced 
a dissipational collapse? In the present study, 
deep $J, H,$ and $K'$ images obtained with the 
ALTAIR adaptive optics (AO) system $+$ NIRI imager on Gemini North (GN) are 
used to probe the age and metallicity distributions in four M31 fields. 
The near-infrared is an important wavelength region for 
studies of this nature. The photometric properties of 
very red stars at visible wavelengths are affected by line blanketing, with the 
result that the upper RGB of stars with near-solar metallicities slumps over on 
visible wavelength CMDs. This makes it difficult to detect the most metal-rich 
upper RGB stars, and so the mean metallicity of the RGB may be underestimated. 
Line blanketing also causes the AGB to be a near horizontal 
sequence on CMDs constructed from data with observations shortward of $1\mu$m 
(e.g. Figure 5 of Richer et al. 1990). However, the AGB forms a sequence that 
is closer to vertical on near-infrared CMDs (e.g. Davidge 2000a), thereby 
making it easier to resolve AGB stars in crowded fields and identify 
the AGB-tip. There is also a well-calibrated relation between the slope 
of the RGB and metallicity in the near-infrared (e.g. Ferraro et al. 
2000; Kuchinski et al. 1995). Finally, interstellar 
extinction is less of a concern in the infrared, 
reducing complications caused by the non-uniform distribution of dust 
throughout the M31 disk (e.g. Cuillandre et al. 2001; Williams 2002).

	The fields that were observed sample a range of stellar densities, from 
the outer disk, where the incidence of crowding is negligible, to a distance of 
only two arcmin from the galaxy center, where crowding is a significant concern 
(e.g. Davidge 2001; Stephens et al. 2001). Three of the fields are dominated 
by bulge stars, while the fourth is dominated by disk stars. With an angular 
resolution of 0.08 arcsec FWHM in $K'$, which corresponds roughly to 0.3 
parsecs at the distance of M31, these are the highest angular resolution 
near-infrared images yet obtained of this galaxy. 

	The paper is structured as follows. Details of the observations, 
the data reduction procedures, and the photometric measurements are discussed 
in \S 2. In \S 3 and 4 the photometric properties of stars 
in the disk and bulge fields are discussed, while in \S 5 the 
issue of crowding is addressed. A summary and discussion of the 
results follows in \S 6.

\section{OBSERVATIONS, REDUCTIONS, \& PHOTOMETRIC MEASUREMENTS}

\subsection{Observations}

	The data were recorded during the night of November 18/19 (UT) 2003 
with the ALTAIR Adaptive Optics (AO) system and NIRI imager on Gemini North 
(GN). ALTAIR is a natural guide star AO system, the key elements of which 
are a 177 element deformable mirror (DM), conjugated to an altitude of 6.5 km 
above the Mauna Kea summit, and a $12 \times 12$ Shack-Hartmann wave front 
sensor (WFS). Herriot et al. (2000) give a complete description of 
ALTAIR. NIRI is the facility infrared imager on GN, and this instrument 
is described in detail by Hodapp et al. (2003).
The f/32 camera was used for these observations, 
and so each pixel on the $1024 \times 1024$ InSb detector 
subtends 0.022 arcsec on a side, while the total imaged field is $22.5 
\times 22.5$ arcsec.

	The target fields were selected based on the availability of a 
guide star with $R < 13$, which is the brightness that will allow ALTAIR 
to provide the maximum correction during typical observing 
conditions. Four fields in M31 that sample regions with very different 
stellar densities were observed through $J, H,$ and $K'$ filters. Table 1 lists 
(1) the field names used throughout this study, (2) the co-ordinates and Guide 
Star Catalogue identification numbers of the guide stars, (3) the angular 
distance from the center of M31, (4) $r-$band disk and bulge surface 
brightnesses for each field based on the Kent (1989) small bulge model, and 
(5) the exposure times for each field. Walterbos \& Kennicutt (1998) compute an 
effective radius of 2.0 kpc for the M31 bulge, which corresponds to 9
arcmin, and 2 of the 4 fields fall within this radius. 
The guide star for Bulge 1 is not in the Guide Star Catalogue 
because of its close proximity to the center of the galaxy.

	A four point square dither pattern, that was 4 arcsec on a side, was 
employed in $J$ and $H$, while a 5 point dither pattern, which was the same 
pattern used in $J$ and $H$ but with an extra point in 
the middle, was used in $K'$. The uncorrected image quality 
while these data were recorded, calculated from data supplied by the 
ALTAIR WFS, ranged between 0.45 and 0.80 arcsec FWHM 
at visible wavelengths. The corrected image quality at a few arcsec 
distance from the AO reference star is typically 0.08 arcsec FWHM in $K'$, and 
0.10 arcsec FWHM in $J$ and $H$.

\subsection{Data Reduction}

	The data were reduced using a standard pipeline for infrared 
observations. The initial steps in the reduction sequence were: 
(1) the subtraction of a dark frame, (2) the division by a flat-field frame, 
(3) the removal of the DC sky level on a frame-by-frame basis, 
and (4) the removal of interference fringes and the thermal signatures of warm 
objects along the optical path. The calibration frame used in the last step was 
constructed by median-combining sky-subtracted images of all M31 fields.

	The processed images were registered to correct for dither offsets 
and then median-combined to suppress bad pixels and cosmic rays. A 
small number of exposures in which the image quality was noticeably poorer than 
average were not included when creating the combined images. Finally, 
the combined images of each field were trimmed to the 
area of common exposure time, which is typically 
$17.5 \times 17.5$ arcsec. The final $K-$band images of the Disk 1 
and Bulge 1 fields, which have the lowest and highest stellar densities 
in the current sample, are shown in Figures 1 and 2.

\subsection{Photometric Measurements}

	Stellar brightnesses were measured with the point spread function (PSF) 
fitting program ALLSTAR (Stetson \& Harris 1988), using PSFs and stellar 
co-ordinates obtained with tasks in DAOPHOT (Stetson 1987). A single spatially 
non-varying PSF, typically obtained from 30 -- 60 stars, was constructed for 
each field $+$ filter combination. Anisoplanicity distorts the PSF near 
the edge of some of our images (e.g. McClure et al. 1991), and only those 
portions of the datasets that were deemed to have stable PSFs, as determined 
from a visual inspection of the sharpness and shape of the PSF, were 
considered for photometric investigation. In $J$ the maximum radius for 
PSF stability was roughly 5 arcsec from the guide star, while in $H$ this 
radius was typically $6 - 7$ arcsec. The PSF was stable over an even 
larger portion of the field in $K$. The Disk 1 data were recorded 
during exceptionally good conditions, and there is no sign of anisoplanicity in 
these data. Consequently, photometry was done over the 
entire field. No attempt was made to do photometry within 2 arcsec of the 
reference star in any field, as the PSF wings of this very bright 
object hinder the detection of faint stars.

	The photometric measurements used in this study were made by TD. 
An independent set of photometric measurements was made by KO, also 
using DAOPHOT. The two sets of calibrated photometry 
agree within the uncertainties, lending confidence to the 
photometric measurements.

	The photometric calibration was set using observations of standard 
stars from Hawarden et al. (2001) and Hunt et al. (1998). The standard 
deviation of the individual standard star measurements 
about the final zeropoints, obtained with a 25 
pixel radius aperture to match the PSF radius used for 
the M31 photometry, is $\pm 0.1$ mag in all three filters. Aperture 
corrections were measured from the PSF stars in each frame. The presence of 
faint, unresolved stars introduces uncertainties in these measurements, that 
will depend on the stellar density of the field in question. We estimate 
that the aperture corrections in Disk 1, Disk 2, and Bulge 2 
have uncertainties of a few hundredths of a mag, while in Bulge 1 
the aperture correction may have an uncertainty of $\pm 0.1$ mag.

	The photometric calibration was checked using the published 2MASS 
brightnesses of the AO guide stars in Bulge 1, Bulge 2, and 
Disk 2; this was not possible for Disk 1 as the AO 
guide star was not detected by 2MASS. The central cores of the guide stars 
are saturated in the images, and so the total brightness was estimated 
from the wings of the PSF. For Bulge 1 and 2 the estimated brightness of the 
guide star agrees with that measured by 2MASS to within a 
few hundredths of a magnitude. However, for 
Disk 2, which has the brightest, and hence most heavily saturated AO 
reference star, there is a 0.25 mag difference in $K$ between the 
brightness measured from the ALTAIR data and that measured by 2MASS, 
in the sense that the ALTAIR brightness is the fainter of the two.
Given that (1) the bright stellar contents in these fields scale well with 
surface brightness (\S 5), and (2) the ALTAIR 
and 2MASS brightnesses of the guide stars in Bulge 1 and 2 are in good 
agreement, we suspect that the AO guide star 
for Disk 2 is a photometric variable.

\subsection{Artificial Star Experiments}

	Artificial star experiments were run to determine sample completeness, 
estimate the uncertainties introduced by photon statistics and crowding, and 
assess systematic effects in the photometry. 
The artificial stars were created using the PSFs constructed for the 
photometric analysis, and so these experiments do not include the effects of 
anisoplanicity. The artificial stars were 
distributed randomly throughout the frames, avoiding the high surface 
brightness region immediately surrounding the AO guide stars. The stars 
were assigned colors ($H-K = 0.3$ and $J-K = 1.2$) and brightnesses ($K$ 
between 16 and 21) appropriate for AGB and RGB stars in M31, and were 
distributed in brightness according to a power-law luminosity function 
(LF) with exponent 0.2. Between 200 and 240 stars were added to each frame over 
two runs, with a minimum of 10 stars per brightness bin. 
This number of stars per run was selected to prevent artificially 
increasing the level of crowding at the faint end in each simulation. While 
the number of artificial stars used in this study is modest, it is 
still sufficient to assess the extent of blending and determine the 
brightness at which sample incompleteness becomes significant in each dataset 
(see below).

	The completeness fractions of stars detected in both $H$ and $K$, 
$C_{HK}$, the mean difference between the actual and measured $K$ brightnesses, 
$\Delta$K, and the standard deviation about $\Delta$K, $\sigma_K$, are 
shown in Figure 3. The difference between the actual and measured 
brightnesses for all of the recovered artificial stars in the two most 
crowded fields, Bulge 1 and 2, is shown as a function of $K$ in Figure 4. 
As $C_{HK}$ drops $\Delta$K and $\sigma_K$ increase, 
and the the magnitude at which sample incompleteness reaches 50\% 
is correlated with field surface brightness, which measures the 
degree of crowding. $\Delta$K is of particular interest as 
it monitors systematic effects in the photometry. In low density environments, 
systematic errors in stellar brightnesses will 
become significant only when statistical errors 
imposed by the night sky level and integration time become large, as faint 
sources that coincide with positive peaks in the sky noise spectrum will 
preferentially be detected. The situation is different for very crowded 
environments, where systematic errors due to crowding, including blending, 
may occur well above the limit imposed by photon statistics. It should be 
noted that in crowded fields the systematic effects on colors, especially 
those spanning small to intermediate wavelength intervals, will likely 
be smaller than in single filter measurements, 
as the photometry in both filters will be 
affected by crowding by roughly the same amount.

	In both disk fields and Bulge 2 $\Delta$K and $\sigma_K$ only depart 
significantly from the values defined at the bright end when $K > 20$, 
although it is evident from Figure 4 that a modest fraction of objects 
in Bulge 2 with $K = 19$ may be blends of fainter objects. That 
$\Delta$K and $\sigma_K$ in these fields are near 0 over much of the 
brightness range explored by the artificial star experiments suggests 
that crowding and blending is not a factor in these fields. 
However, the situation is different in Bulge 1, which is the most 
crowded field. In this field $\Delta$K $< 0.1$ magnitude 
when $K < 18.5$, while $\sigma_K$ grows quickly when $K > 19$. 

	A modest fraction of objects in Bulge 1 at the bright end are likely 
blends. Indeed, of the 45 artificial stars with $K \leq 18$, 3 were recovered 
with brightnesses that exceed the actual values by at least 0.5 magnitudes, 
indicating that they were blended with stars of similar brightness.
The frequency of such blends is thus $3/45 = 0.07^{+0.06}_{-0.02}$, where the 
quoted errors are $1-\sigma$ uncertainties. 

\section{RESULTS: DISK 1}

	Disk 1 differs from the other fields observed with ALTAIR 
in that (1) the stellar density is low, so that crowding is not an issue, 
and (2) the stellar content is dominated by disk stars. 
The $(K, J-K)$ and $(K, H-K)$ CMDs of stars in Disk 1 
are plotted in the left hand panels of Figures 5 and 6. The number of stars 
plotted in each CMD is stated in each panel. Given the modest 
number of bright giants detected in this field then only limited conclusions 
can be drawn from these data.

	The RGB in Disk 1 is wider than what is 
expected soley from the photometric errors predicted from the artificial star 
experiments. This is demonstrated in Table 2, where the standard 
deviation in $H-K$ and $J-K$ for stars with $K$ between 17.5 and 18.5, 
which corresponds roughly to the top 1 magnitude of the RGB, listed in columns 
2 and 4, are compared with the dispersion predicted from the 
artificial star experiments, which are listed in columns 3 and 5. An 
F-test indicates the observed and predicted dispersions differ 
well in excess of the $2-\sigma$ significance level.

	A number of factors likely contribute to the broadening of the RGB, 
including anisoplanicity, differential reddening, 
and the dispersion in mean metallicity that is seen throughout the 
disk and halo of M31 (Bellazzini et al. 2003). 
The latter may dominate. In fact, the metallicity distribution functions 
(MDF) constructed by Bellazzini et al. (2003) have an approximate 
standard deviation of $\pm 0.5$ dex in [Fe/H], and this will introduce a 
dispersion of $\pm 0.1$ mag in $J-K$ near the RGB-tip (Ferraro et al. 2000). 
The result of subtracting in quadrature the scatter due to photometric 
errors from the observed dispersion in $J-K$ is shown in the last column 
of Table 2. This `residual' scatter is close to that predicted from a 
$\pm 0.5$ dex dispersion in metallicity.

	While the $(K, H-K)$ CMD goes considerably deeper than the $(K, J-K)$ 
CMD, $J-K$ is much more sensitive to metallicity than $H-K$, and so the $(K, 
J-K)$ CMD is used to investigate the metallicity of stars in this study. 
The left hand panel of Figure 7 shows the $(M_K, (J-K)_0)$ CMD of 
Disk 1, assuming a distance modulus of 24.4 
(van den Bergh 2000), and E(B--V) = 0.1 due to foreground material 
(Burstein \& Heiles 1984). These values are used throughout 
the paper. The reddening law from Rieke \& Lebofsky (1985) was adopted 
to compute A$_K$ and E(J--K). Also shown on Figure 7 are the RGB loci of the 
globular clusters 47 Tuc ([Fe/H] = --0.8) and NGC 6528 ([Fe/H] 
= 0.0), based on the RGB colors listed in Table 2 of 
Ferraro et al. (2000); the cluster metallicities are taken from the Harris 
(1996) database. 

	Inspection of Figure 7 suggests that Disk 1 is dominated by 
relatively metal-rich stars, with a mean [Fe/H] close 
to that of NGC 6528. The MDF is evidently skewed 
to high values, as no stars significantly more metal-poor than 47 Tuc 
have been detected, although with only a 
modest number of stars in the CMD then even a population 
of giants that are more metal-poor than 47 Tuc and account for $\leq$ 10\% 
of the total number of stars may be missed due to small 
number statistics. Nevertheless, the relatively high mean metallicity 
of the Disk 1 field is a robust result, that is consistent with other 
studies of the outer regions of M31 (e.g. Bellazzini et al. 2003).

	The $K$ LF of Disk 1, based on stars that are 
detected in both $H$ and $K$, is shown in Figure 8. 
Whereas a 0.25 mag binning interval is used to construct the LFs of the 
fields closer to the center of M31, the low density of stars 
in Disk 1 necessitates the use of a coarser binning interval. 
Despite the large error bars in the individual bins, it is evident 
that the Disk 1 LF roughly follows a power-law.

	Davidge (2000b) used the CFHT AO system to investigate the stellar 
content in the metal-rich globular cluster NGC 6528, which has a metallicity 
and age that is comparable to that of stars in Baade's Window (e.g. 
Momany et al. 2003). A least squares fit of a power-law to the NGC 6528 
LF shown in Figure 3 of Davidge (2000b), neglecting the bins containing the 
HB and RGB-bump, gives an exponent $x = 0.21$. For comparison, Davidge (2001) 
found that the $K$ LFs of metal-poor globular clusters tend to have 
$x=0.3$, although there is considerable scatter, 
with some clusters having exponents similar to that measured in 
NGC 6528. Given that the RGB in Disk 1 has a color consistent 
with NGC 6528, a power-law with $x = 0.21$ was fit to the LF entries with 
$K$ between 18 and 21, and the result is plotted in Figure 8. 
The $x = 0.21$ power-law provides a reasonable fit to the data, 
although there is considerable scatter due to small number statistics. 

	We have not attempted to measure the brightness of the RGB-tip in 
Disk 1 because of the modest number of stars. However, it is evident from 
Figure 7 that the vast majority of the stars in Disk 1 have 
M$_{bol} > -4$, and hence are on the RGB (Ferraro et al. 2000). 
There is only one star significantly brighter than the RGB-tip in 
Disk 1, and this object is likely evolving on the AGB.

	An intermediate age population might be expected in 
Disk 1 given the detection of intermediate age stars in the 
halo (Brown et al. 2003) and outermost disk (Ferguson \& Johnson 2001) of 
M31. Unfortunately, the AGB content of Disk 1 provides only loose 
limits on age because of the tiny area sampled by ALTAIR$+$NIRI coupled 
with the low surface density of stars. The single bright AGB 
star has M$_K = -7.5$. Models predict that the AGB-tip 
should occur near M$_K = -7.5$ in solar metallicity systems with ages in 
excess of $\sim 5$ Gyr (Girardi et al. 2002), and this is consistent with the 
intrinsic $K-$band brightnesses of long period variables (LPVs) in metal-rich 
globular clusters (e.g. Frogel 1983). Therefore, based solely on the 
single bright AGB star detected here, there is no 
evidence for a population younger than a few Gyr, 
although this conclusion may change when a larger area is surveyed.

\section{RESULTS: THE INNER DISK AND BULGE}

\subsection{Disk 2}

	The $(K, H-K)$ and $(K, J-K)$ CMDs of Disk 2 are shown 
in the right hand panels of Figures 5 and 6, while the $(M_K, (J-K)_0)$  CMD 
is shown in the right hand panel of Figure 7. The standard deviations of the 
colors of stars with $K$ between 17.5 and 18.5 are listed in Table 2. As 
in Disk 1, the Disk 2 RGB is wider than 
expected from photometric errors alone. The scatter that remains 
after removing the contribution from photometric errors is broader than in 
Disk 1.

	The brightest AGB stars in Disk 2 have M$_K$ between 
--8 and --8.5, and the most luminous star has M$_{bol} = -5$. The ridgeline of 
the RGB in Disk 2 is 0.2 magnitude in $J-K$ redder than in the 
Disk 1. This difference exceeds the 0.1 magnitude uncertainty in the 
photometric calibration, and so suggests a higher 
mean metallicity for Disk 2, as might be expected given the evidence for a 
radial abundance gradient in the M31 disk (e.g. Blair, Kirshner, \& 
Chevalier 1982). However, Disk 2 is on the eastern boundary 
of dust lane D517 (Hodge 1981), and so reddening internal to M31 may 
contribute to the red RGB color. The amount of internal reddening 
could be significant, as globular clusters within a tenth of a degree of 
Disk 2 have E(B--V) as high as 0.4 (Barmby et al. 2000), which 
corresponds to $E(J-K) \sim 0.1$ mag. Correcting for an internal 
reddening contribution of this size would bring the RGB locus of 
Disk 2 in Figure 7 into better agreement with that of Disk 1.

	The $(J-H, H-K)$ two-color diagram (TCD) of stars in Disk 2 
with $K < 18$, which includes the AGB and upper RGB, is shown in Figure 
9. The stars form a sequence on the TCD that overlaps the locus of solar 
neighborhood giants and LPVs in the Magellanic Clouds. 
The spectral energy distributions (SEDs) of the 
brightest stars in Disk 2 appear to differ from those in 
Disk 1, which are also plotted in Figure 9, in the sense that the stars 
with the reddest $H-K$ colors in Disk 1 have smaller $J-H$ 
colors than stars in Disk 2. The near-infrared SEDs of 
the brightest stars in Disk 1 are similar to the brightest 
variables in the old moderately metal-poor globular cluster 47 Tuc, which 
were investigated at near-infrared wavelengths by Frogel, Persson, \& Cohen 
(1981).

	The $K$ LF of Disk 2, constructed from stars detected in 
both $H$ and $K$, is shown in the lower panel of Figure 8. The LF follows a 
power-law at the faint end, and the result of fitting a power law with exponent 
0.21, based on the LF of RGB stars in the globular cluster NGC 6528 (\S 3), 
to the data between $K = 18$ and 21 is shown as a dashed line. The 
fitted relation provides a reasonable match to the data. There is 
a discontinuity near $K = 17.75$, which corresponds 
to M$_K = -6.6$ with the adopted distance and 
foreground reddening, that we identify as the RGB-tip. This RGB-tip 
brightness is $0.1 - 0.2$ mag fainter than what is seen in the 
most metal-rich globular clusters studied by Ferraro et al. (2000); 
consequently, correcting for internal extinction of the size discussed earlier 
would bring the measured RGB-tip brightness into agreement with what is seen 
in the most metal-rich globular clusters. 

	Stars on the AGB evolve at a pace that is roughly four times 
faster than stars on the RGB, and in the lower panel of Figure 8 the power-law 
that was fit to the Disk 2 data has been shifted down the vertical 
axis by an amount that accounts for this faster evolution. The shifted 
relation matches the data with $K < 17.75$ within 
the estimated uncertainties, although the data points 
fall systematically above the expected AGB sequence.

	There is a population of bright stars with $(H-K) > 0.6$ that form a 
loose sequence that departs from the giant branch in the $(K, H-K)$ CMD. 
Objects with similar colors were detected by Stephens et al. (2003) and were 
classified as long period variables (LPVs) in that study. Luminous 
LPVs in the LMC with $(H-K) > 0.6$ and $(J-K) > 1.6$ tend almost exclusively 
to be C stars (Hughes \& Wood 1990), and so we identify the very red 
objects in Disk 2 as C stars. We note that there may be 
a mild metallicity sensitivity to the critical color for identifying C stars, 
as only 4 of the 6 spectroscopically confirmed C stars above the RGB-tip in 
the SMC, which is roughly a factor of two more metal-poor than the LMC, 
listed in Table 1 of Wood, Bessell, \& Fox (1983) have $(J-K)_0 > 1.6$, 
although one of the two `missed' stars has $(J-K)_0 = 1.59$.  

	The detection of bright C stars is
direct evidence of an intermediate age population in Disk 2. 
In fact, C stars contribute a significant fraction of the total AGB luminosity 
in Disk 2. To estimate the fractional contribution made by C 
stars to the total light output of stars above the RGB-tip, bolometric 
luminosities were computed for stars with $K < 17.5$ using 
the relation between K-band bolometric correction and $J-K$ from Bessell \& 
Wood (1984). Using this procedure, we find that the 6 C star candidates in 
Disk 2 account for $29^{+18}_{-8}$ \% of the 
total AGB luminosity. This result is sensitive to the 
threshold color used to define C stars, and the uncertainty in the 
ratio quoted above was computed by changing the $H-K$ color threshold 
by $\pm 0.1$ magnitude, which is the uncertainty in the photometric 
calibration. The sizeable uncertainties in the contributions made by C stars 
to the total AGB light notwithstanding, it appears that C stars in Disk 2 
contribute a larger fraction of the total AGB light than near the 
center of NGC 205, where C stars account for roughly 10\% of the total AGB 
luminosity (Davidge 2003b). 

\subsection{Bulge 1 and 2}

	The $(K, H-K)$ and $(K, J-K)$ CMDs of the fields dominated by bulge 
stars are shown in Figures 10 and 11. As can be seen from the entries in Table 
2, the width of the RGB in Bulge 2 is comparable to what is seen in 
Disk 1 and 2. While the giant branch in 
the $(K, J-K)$ CMD of Bulge 1 is broader than in any of the other three 
fields, due to the larger uncertainties in the photometry arising from 
the high stellar density, the residual scatter after accounting for the 
photometric errors is consistent with what is seen in the other fields.

	The $(M_K, (J-K)_0)$ CMDs of the bulge fields are shown in Figure 
12. All of the stars in Bulge 2 have M$_{bol} > -5$, whereas there are 
three stars brighter than M$_{bol} = -5$ in Bulge 1. After adjusting for 
foreground reddening and the distance to M31, the ridgeline of the 
Bulge 1 RGB coincides with the RGB of NGC 6528. For comparison, the 
Bulge 2 giant branch falls 0.2 magnitudes in $J-K$ redward of the NGC 6528 
sequence, which is an amount that is greater than the uncertainty in the 
photometric calibration. Some of the most heavily reddened globular clusters in 
M31 are seen in the central regions of the galaxy (Figure 8 of Barmby et al. 
2000), raising the possibility that there may be considerable dust 
along inner bulge sight lines. If, as indicated by globular clusters in the 
Barmby et al. (2000) sample, $E(B-V) > 0.4$ then the intrinsic 
$J-K$ colors of the RGB in some regions near the center of M31 may 
be $0.1 - 0.2$ mag bluer in $J-K$ than observed. If internal 
extinction of this size is present in Bulge 2 then the RGB of this field would 
have an intrinsic color that agrees with the NGC 6528 sequence. 

	The $K$ LFs of Bulge 1 and 2, based on stars detected in both $H$ and 
$K$, are compared in Figure 13. The dashed lines are power laws with exponent 
$x=0.21$, computed from the LF of RGB stars in the globular cluster NGC 
6528 (\S 3), that were fitted to the LF entries between $K = 18$ and 20. 
The dashed-dotted line is the fitted sequence shifted down by 80\% to show 
the approximate relation for an AGB population. The power-law fitted to the RGB 
provides a reasonable match to the data at the faint end. A 
discontinuity is seen in the Bulge 2 LF between $K = 17.25$ and 17.50, and 
the amplitude of the discontinuity is consistent with a 
transition from a population dominated by first ascent giants to one 
dominated by AGB stars. As for Bulge 1, the LF departs from the RGB trend near 
$K = 17.0$, suggesting that the RGB-tip in this field is brighter than 
in Bulge 2, although we caution that the artificial star experiments 
indicate that a modest fraction of the stars at this brightness may be 
blends. The relative number of stars brighter 
than the RGB-tip is consistent with a transition from 
RGB to AGB dominated populations. In both fields there is a 
bin near the RGB-tip with star counts that are intermediate between 
the RGB and AGB sequences, indicating that binning errors may introduce 
a $\sim 0.1$ mag uncertainty in the RGB-tip measurement. Nevertheless, the 
RGB-tip brightnesses inferred from the discontinuities in the LFs of both 
fields correspond to M$_K$ between --7 and --7.5, with the fainter of these 
being comparable to the brightness expected for an old, solar metallicity 
population (Ferraro et al. 2000).

	The near-infrared SEDs of sources in Bulge 1 and 2 
are investigated in Figure 14, which shows the $(J-H, H-K)$ TCD of stars with 
$K < 18$ in both fields. As might be expected from the CMDs, the near-infrared 
SEDs of the majority of sources in both fields are consistent with them 
being late M-type giants. There is also a spray of stars having 
SEDs that tend to fall below the Magellanic Cloud LPV sequence, in a part of 
the TCD that is occupied by bright variable stars in the Galactic bulge 
(Frogel \& Whitford 1987). 

	The color distributions of AGB stars in Bulge 1 and 2 appear to 
differ, in the sense that red stars dominate when $M_K < -7.5$ in Bulge 1, 
while in Bulge 2 the majority of stars with $M_K < -7.5$ have 
$J-K \sim 1.4 \pm 0.1$, and are likely oxygen-rich M giants. However, 
these differences are not statistically significant. More importantly, 
both fields contain some stars with $K < 17$ that have $H-K > 0.6$ and $J-K 
> 1.6$, which we identify as C stars (\S 4.1).

	The contribution that C stars make to the total 
AGB luminosity in Bulge 1 and 2 is comparable to that 
in Disk 2. After computing bolometric corrections 
using the procedure discussed in \S 4.1, we find that the 6 C star 
candidates in Bulge 1 account for $17^{+18}_{-3}$ 
\% of the total AGB light, while the 2 C stars in Bulge 2 
contribute $7^{+15}_{-2}$ \% of the total AGB luminosity.
As in \S 4.1, the uncertainties reflect the effect of 
changing the color threshold for identifying C stars by $\pm 0.1$ mag. 
Stephens et al. (2003) argued that the number density of very red stars, 
when normalized to bluer stars, does not change from 
field-to-field. The ALTAIR data support this conclusion, although 
we emphasize that the uncertainties in the fractional C star luminosities are 
substantial.

\section{CROWDING}

\subsection{The Incidence of Blending in the ALTAIR Data}

	Blending occurs when two or more stars fall within the same 
angular resolution element, and so are identified as a single object. In a very 
crowded field such as Bulge 1 it is likely that all resolution elements 
contain more than one star, but the effect of a very faint source blending 
with a very bright source has only a minor impact on the photometric properties 
of the latter. The most common significant blending event involves two stars 
with comparable brightnesses, in which case an object will be produced that is 
roughly 0.6 mag brighter than the progenitors, and the artificial star 
experiments discussed in \S 2.4 indicate that systematic offsets of this 
size do not become a factor until well below the RGB-tip in the two 
disk fields and Bulge 2. Blends do occur at brighter magnitudes in 
Bulge 1, but the majority of stars when $K < 18$ are unblended objects.

	Stephens et al. (2003) used simulations to investigate the effects of 
blending in HST NICMOS observations of various fields in M31, and these 
simulations provide a check on our artificial star experiments. Stephens et al. 
(2003) found that crowding is not a factor among the brightest stars in the 
NICMOS data when the $K-$band surface brightness 
$> 16$ mag arcsec$^{-2}$, which corresponds 
roughly to an $r-$band surface brightness $> 18.9$ mag arcsec$^{-2}$. 
When the $K-$band surface brightness $< 16$ mag arcsec$^{-2}$ then 
a significant fraction of the sources detected with NICMOS above the RGB-tip 
may be blends, and the peak stellar brightness could be elevated by 
0.5 mag in $K$. 

	The ALTAIR observations have an angular resolution 
in $K$ that is over a factor of two better than that delivered by 
NICMOS, and so the incidence of blending
will be reduced with respect to the Stephens et al. (2003) data. 
Indeed, the ratio of the areas of the resolution elements in the ALTAIR and 
NICMOS data is $(0.08/0.17)^2 = 0.22$. Therefore, based on the Stephens et al. 
(2003) simulations, blending among the brightest stars should not 
significantly affect the conclusions drawn from these data until 
the $r-$band surface brightness $< 17.3$ mag arcsec$^{-2}$; 
for comparison, the surface brightness in Bulge 1 is 18.1 mag arcsec$^{-2}$. 
This expectation is consistent with the results from the artificial 
star experiments for Bulge 1 discussed in \S 2.4, which indicate that the 
vast majority of stars with $K < 17$ are not blends.

\subsection{Comparisons with the Disk 1 and HST NICMOS Data}

	Comparisons between the four fields observed with ALTAIR provide 
direct checks on the extent of crowding. Disk 1 is of particular 
interest as crowding is not an issue in this part of M31. If it is 
assumed that the stellar contents in all four fields are similar then, if the 
upper RGB content is not affected by crowding, the number density of stars 
near the RGB-tip should scale with surface brightness. This issue is 
investigated in Figure 15, where the $K-$band LFs of Disk 2 and 
both Bulge fields are compared with the LF of Disk 1. The four fields 
have very different stellar densities, and the number counts of the fields 
close to the center of M31 were scaled down to match the stellar density in 
Disk 1 using Kent's (1987) $r-$band surface brightness measurements. 
When normalized in this manner the number counts in Disk 2 and Bulge 2 
agree with the Disk 1 data, indicating that 
blending does not affect the data for these fields. As for Bulge 1, 
the LF of this field consistently falls above that of Disk 1, 
although in most instances the difference 
is significant at only the $1-\sigma$ level.

	Additional insights into crowding in the ALTAIR 
observations can be gleaned from comparisons with other datasets. 
The NICMOS data from Stephens et al. (2003) are of particular interest, 
as these have an angular resolution approaching that of the 
ALTAIR data, and sample fields spanning a range of stellar densities.
In fact, despite significant differences in angular resolution and pixel 
sampling, we find the same peak brightnesses for the RGB 
as measured from the NICMOS data. In particular, 
Stephens et al. (2003) find that the RGB-tip 
occurs near M$_K = -7$, with $(J-K)_0 \sim 1.2$, and this is roughly 
consistent with what is seen in the ALTAIR observations. Stephens et al. (2003) 
also find that the brightest stars have M$_K$ between --8 and --8.5, 
and this peak brightness is similar to what is seen here.

	A more rigorous test of crowding is to compare the number 
densities of fainter AGB stars and stars on the RGB, 
where the incidence of blending is expected to be greater than on the 
upper AGB, in the Stephens et al. (2003) and ALTAIR 
data. This is done in Figure 16, where the $K-$band LFs of Bulge 1 and 2 
are compared with the NIC2 $K-$band LFs of fields F1, F174, F177, which 
have the highest stellar densities examined by Stephens et al. (2003). 
As in previous comparisons, differences in stellar density 
have been removed using the Kent (1987) $r-$band surface 
brightness measurements. When scaled in this manner the stellar contents of 
F1, F174, and F177 are in good agreement, although there is scatter amounting 
to a few tenths of a dex above the RGB-tip. The surface brightnesses of F1 and 
F174 differ by almost 1 magnitude per arcsec$^{2}$, and the excellent agreement 
between the LFs of these fields in Figure 16 suggests strongely that blending 
is not a major factor at the bright end of the NICMOS data.

	The Bulge 2 LF is in good agreement with the NICMOS measurements 
over most of the brightness range considered. While the Bulge 1 data 
consistently falls some 0.3 -- 0.4 dex above the LFs of the NICMOS fields, 
this is likely not a consequence of crowding. The artificial star experiments 
indicate that blends may account for $\sim 10\%$ of sources in Bulge 1 at $K = 
17$, and the offset between the LFs of Bulge 1 and the other fields in 
Figure 16 is greater than this. More importantly, F1 has a higher surface 
brightness than Bulge 1, and it was observed with a coarser angular resolution. 
Therefore, if blending is a factor in any of the datasets shown in Figure 16
then it will be F1 that will be most affected, and this is clearly 
not the case. It is also worth noting that the Bulge 1 and the 
Stephens et al. (2003) data are in good agreement at the faint end, which 
is where differences due to blending should be most apparent.
We suspect that the elevated star counts at the bright end of Bulge 1 
may be due to spatial flucuations in the stellar content along the line of 
sight.

	In summary, the comparisons in Figures 15 and 16 indicate that the 
majority of objects in the ALTAIR data are not affected by crowding. 

\section{DISCUSSION \& SUMMARY}

	Near-infrared images with angular resolutions approaching the 
diffraction limit of the 8 meter GN telescope have been used to investigate the 
photometric characteristics of bright evolved stars in the bulge and inner disk 
of M31. These are the highest angular resolution images yet obtained of 
M31 at these wavelengths, and the photometric measurements of the brightest 
red stars are not affected by blending, even in the field that is closest to 
the center of the galaxy (\S 5). Thus, these data provide unique insights 
into the stellar content in the central few arcmin of M31.

\subsection{Stellar Content Trends in the Disk and Bulge of M31}

	Bellazzini et al. (2003) found a remarkable similarity in the 
metallicity properties of RGB stars in a number of M31 fields that (1) 
sample a range of galactocentric radii, and (2) are well outside of the bulge. 
The data discussed in the present paper extend field-to-field comparisons 
of RGB properties to smaller, bulge-dominated, galactocentric radii. 
Indeed, the galactocentric radius of Bulge 1 is an order of magnitude 
smaller than the field closest to the center of M31 studied by Bellazzini et 
al. (2003).

	In \S 3 and 4 it was demonstrated that the RGBs of Disk 1 and 
Bulge 1 have $J-K$ colors that are consistent with the RGB of the metal-rich 
globular cluster NGC 6528, while the RGBs of Disk 2 and Bulge 2 have 
$J-K$ colors that are redder than the NGC 6528 sequence. Before discussing 
this result further, it should be noted that 
other colors give similar results. In particular, we have compared 
the $(H, J-H)$ CMDs of the four fields observed with ALTAIR with 
fiducial globular cluster sequences from Table 6 of 
Valenti, Ferraro, \& Origlia (2004). The RGBs of Disk 1 and Bulge 1 on the 
$(H, J-H)$ CMDs agree well with the RGB of the 
metal-rich globular cluster NGC 6440, which has a metallicity that is 
only slightly lower than that of NGC 6528, while the RGBs of Disk 2 and 
Bulge 2 are significantly redder. 

	The mean metallicities of the four fields observed with ALTAIR, as 
inferred from the color of the RGB, are likely reliable to no more than a 
few tenths of a dex.  The uncertainty in the photometric calibration is $\pm 
0.1$ magnitude, which introduces uncertainties of $\pm 0.3$ dex in 
[Fe/H]. An even greater source of potential error is the patchy dust 
absorption that occurs throughout the central 
regions of M31. In \S 4.2 it was argued that this introduces 
uncertainties of $0.1 - 0.2$ magnitudes in the intrinsic RGB colors, 
with the result that the mean metallicity in each field can not be 
estimated to better than a few tenths of a dex. 

	While the color of the RGB can be affected by uncertainties in the 
photometric calibration and the mean line-of-sight extinction, this is not 
the case for the scatter at a given brightness within a particular field. 
After accounting for the dispersion caused by photometric errors, as predicted 
by artificial star experiments, the widths of the upper RGBs 
of all four fields on the $(K, J-K)$ CMDs are consistent with a metallicity 
dispersion with a standard deviation $\pm 0.5$ dex in [Fe/H]. This is 
similar to the dispersion detected by Bellazzini et al. (2003) in a large 
number of fields throughout the disk and halo of M31. 

	The similarity in stellar content between the four fields observed 
with ALTAIR is not restricted to metallicity and metallicity dispersion. 
In \S 5.2 it was demonstrated that the $K-$band 
LFs of the 4 ALTAIR fields scale with $r-$band surface brightness. In old 
solar metallicity populations the integrated $r-$band light is dominated by 
main sequence stars (e.g. Buzzoni 1989). For comparison, 
the $K-$band LFs constructed from the ALTAIR data 
are dominated by stars on the upper RGB and AGB. The 
comparisons in Figure 15 thus show that the ratio of AGB $+$ RGB stars to 
main sequence stars, which is a quantity that is sensitive to age, 
appears not to change over a range of disk-to-bulge ratios.

	In the following sub-sections the results 
obtained from the ALTAIR data are discussed in the context of other 
observations. Given that disks and bulges are 
structurally distinct entities, the stellar contents of the disk and bulge 
of M31 are discussed separately. However, the reader is reminded that there are 
mechanisms that may couple the stellar contents of disks and bulges 
(e.g. Kim \& Morris 2001).

\subsubsection{The Disk}

	That the metallicities of RGB stars in Disk 1 and 2 differ by no more 
than a few tenths of a dex suggests that the metallicity gradient among RGB 
stars in the disk of M31 must be modest in size, and this conclusion is 
consistent with some, but not all, studies of the chemical content in M31. 
For example, Trundle et al. (2002) find that the chemical contents 
of bright supergiants do not show radial variations in M31, while 
Bellazzini et al. (2003) find that the abundance distributions of RGB stars 
in a number of fields are similar. In contrast, 
Blair, Kirshner, \& Chevalier (1982) find radial gradients in the abundances 
of nitrogen and oxygen in M31 HII regions, which are consistent with 
radial variations in the C to M giant ratio when R$_{GC} < 20$ kpc if these 
are interpreted as a metallicity effect (Brewer, Richer, \& Crabtree 1995).

	Radial variations in mean age and metallicity are common 
in spiral galaxies (e.g. Bell \& de Jong 2000), and 
gradients of this nature were likely imprinted during disk formation. However, 
as noted in the previous paragraph, there are some indications 
that mixing may have been highly effective throughout a 
large portion of the M31 disk, at least among stars with ages comparable to 
objects on the RGB. What processes could have smoothed population gradients in 
the disk of M31? 

	Radial flows of gas and stars in disks can be induced 
by a number of processes, including spiral density 
waves (e.g. Casuso \& Berkman 2001; Sellwood \& Binney 2002), 
viscosity (Silk 2001; Ferguson \& Clarke 2001), galaxy-galaxy interactions 
(Barnes \& Hernquist 1992; Mihos \& Hernquist 1996), dynamical friction 
(Noguchi 1999; 2000) and the infall of angular momentum-rich gas from the 
halo (Ferguson \& Clarke 2001). The outward movement of gas clouds may 
drive the age distribution near the disk edge to younger values (Ferguson \& 
Clarke 2001). In fact, intermediate age populations at large 
galactocentric distances, possibly related to the disk, are 
seen in the Sc galaxies NGC 2403 and M33 (Davidge 2003a), as well 
as in the outer regions of the M31 disk (Ferguson \& Johnson 2001). 

	Of the various mechanisms that might cause mixing, tidal interactions 
between M31 and its companions are potentially attractive given the evidence 
that M31 and its satellites have interacted. For example, there is a warp in 
the HI disk (e.g. Sawa \& Sofue 1982), while the star forming history of NGC 
205 appears to be coupled with its orbit about 
M31 (Davidge 2003b). Perhaps most significantly, there are 
well-defined tidal tails extending from some M31 companions (e.g. Ferguson 
et al. 2002). However, tidal interactions should also cause mixing 
of gas, and so this process does not explain the abundance gradients found by 
Blair et al. (1982).  Clearly, it will be of interest to determine if 
RGB stars in other disk systems, such as M81, show evidence for an abundance 
gradient, or are as well mixed as in M31.

\subsubsection{The Bulge}

	The bulge of M31 likely contains a radial metallicity gradient. Davidge 
(1997) estimated that $\Delta$[Fe/H]/$\Delta$log(r) $= -0.5$ from a grid 
of long slit spectra that mapped the central 1 arcmin of M31. This 
is shallower than what is seen in the outer regions of the bulge of the 
Milky-Way (Minniti et al. 1995; Tiede, Frogel, \& Terndrup 1995), although 
Ramirez et al. (2000) find that the metallicity gradient in the Galactic 
bulge may flatten in the central few hundred parsecs. 

	If the gradient measured by Davidge (1997) is assumed to extend to 
larger radii then the mean metallicities of Bulge 1 and 2 will differ by 0.15 
dex. This is likely a lower limit as the abundance gradient outside of the 
inner regions of the M31 bulge may steepen, as is the case in the Milky-Way 
(Ramirez et al. 2000). The Bulge 1 and 2 data may contain signs of a 
metallicity gradient in the M31 bulge. In \S 4.2 it was demonstrated that the 
RGB-tip in Bulge 1 appears to the brighter than that in Bulge 2. While 
the artificial star experiments indicate that some of the stars in Bulge 
1 near the RGB-tip are likely blends, the number of such objects will 
account for only $\sim 7\%$ of the total at this 
brightness (\S 2.4), and this is clearly not enough to explain the difference 
in the RGB-tips in Figure 13. We further note that the 
difference in RGB-tip brightness is also not a consequence of small 
number statistics, as numerical simulations indicate 
that only 45 stars are needed on the upper 2.5 mag of the 
RGB to measure the RGB-tip brightness to within $\pm 0.1$ mag (Crocker \& Rood 
1984), and the upper RGBs of Bulge 1 and 2 contain far more stars than this.

	A metal-rich population in Bulge 1 that is not seen in Bulge 2 could 
explain the difference in RGB-tip brightnesses, as the brightness of the 
RGB-tip in $K$ depends on metallicity, in the sense of becoming brighter and 
redder with increasing metallicity. The calibration between the $K-$band 
brightness and metallicity among globular clusters derived by 
Ferraro et al. (2000) indicates that a 0.15 dex difference in 
metallicity would change the RGB-tip brightness by roughly 0.1 magnitude, 
and a shift of this nature would bring the Bulge 1 and 2 LFs in Figure 13 
into better agreement. It should be emphasized that the RGB colors of 
Bulge 1 and 2 are not consistent with this interpretation, although this could 
be a consequence of reddening internal to M31 (e.g. \S 4.2). 
Clearly, spectroscopic observations of individual stars will be the most secure 
means of measuring metallicities, and integral field spectrographs coupled to 
adaptive optics systems, such as ALTAIR $+$ NIFS on Gemini North, will be 
required to overcome crowding in the high stellar density fields near the 
center of M31. 

\subsection{The Bright AGB Content Near the Center of M31 and the Presence of C Stars}

	One of the goals of photometric surveys of nearby spiral galaxies is 
to probe the star-forming histories of their pressure and rotationally 
supported components, as this will provide insight into when their 
global morphological characterics were imprinted. In the case of the 
Milky-Way, the relative age of the disk and bulge is still a matter 
of debate. Studies of the main sequence turn-off in the Galactic bulge
indicate that it has an age that is similar to that of the majority of globular 
clusters (Feltzing \& Gilmore 2000; Kuijken \& Rich 2002; Zoccali et al. 2003). 
However, the age of the Galactic disk is less certain. Binney, Dehnen, \& 
Bertelli (2000) conclude that the solar neighborhood contains stars 
that are at least as old as the oldest globular clusters, although 
studies of white dwarfs suggest an age of only 8 Gyr for the 
solar disk (Leggett, Ruiz, \& Bergeron 1998, but see also Hansen 1999).

	Peak AGB brightness offers one means of investigating the age of stars 
near the center of M31, although there are significant problems with this 
statistic as an age indicator. First, the three fields observed with ALTAIR 
near the center of M31 contain a mixture of bulge and disk stars. 
Lacking kinematic information for individual objects, the task of 
isolating pure samples of disk and bulge stars is problematic. Second, the 
brightness of the AGB peak depends on metallicity as well as age, so that 
there is an age-metallicity degeneracy when interpreting the peak brightness. 
A young or intermediate-age component that is moderately metal-poor may 
also be missed if there is an older but more metal-rich population with a 
brighter AGB-tip. Third, efforts to measure the peak brightness of the AGB are 
complicated by stellar variability. If the brightest AGB stars in the bulge 
of M31 are LPVs like those in the Galactic bulge then they will have 
amplitudes of up to $\pm 1.0$ mag in $K$ (Glass et al. 1995). In fact, it is 
likely that the brightest stars imaged in the bulge of M31 at any given 
time are LPVs near the peak of their light curves, and this can be confirmed 
with multi-epoch observations, as has been done for M32 by 
Davidge \& Rigaut (2004). 

	The brightest AGB stars in Bulge 1 and 2 have M$_K < -8.5$, and 
M$_{bol} = -5$, and hence are not different from what is seen in 
the Galactic bulge (Frogel \& Whitford 1987). If these stars are LPVs 
near the peak of their light curves then they will have a mean brightness 
M$_K \sim -7.5$. This is consistent with the expected AGB peak brightness 
in an old solar-metallicity system (Girardi et al. 2002). The peak AGB 
brightness in the M31 bulge is also comparable to that of the brightest AGB 
star in the metal-rich globular cluster NGC 6553 (Guarnieri, Renzini, \& 
Ortolani 1997), which has an old age (e.g. Ortolani et al. 1995). These data 
thus suggest that an old population dominates in the inner regions of M31. 

	As noted above, peak stellar brightness does not offer an ironclad 
means of probing age. In fact, there are a number of sources with M$_K < -8$ in 
Bulge 1 and 2 and the Disk 2 field that have very red colors, and these 
provide additional insights into the stellar content near the 
center of M31 that contradicts the notion that these fields are dominated 
by an old stellar population. In particular, Hughes \& Wood (1991) investigated 
the near-infrared photometric properties of LPVs in the LMC, and found that the 
vast majority of objects with $J-K > 1.6$ are C stars; of the three objects 
in their sample with $J-K > 1.6$ that are not C stars, two are very luminous 
(M$_{bol} < -5.5$) and have periods in excess of 600 days. Davidge (2003b) 
used $H-K$ and $J-K$ colors to identify C stars in NGC 205, and 
recovered C stars identified previously by Richer, 
Crabtree, \& Pritchet (1984) from narrow-band photometry.

	Based on the photometric study by Hughes \& Wood (1991), we identify 
objects with $J-K > 1.6$ in the ALTAIR CMDs as C stars. These stars tend to 
have M$_K$ between --7 and --8.5. Observations indicate that C star 
production does not occur in systems with ages in excess of 6 Gyr, and that the 
maximum age for C star production drops as one moves to higher metallicities 
(e.g. Cole \& Weinberg 2002). Hence, the presence of C stars indicate that 
there is likely an intermediate age population in Disk 2 and Bulge 1 and 2.

	The identification of the very red objects in M31 as C stars has 
potential implications for the bulge and inner disk of the Milky-Way, as a 
population of very red objects is also seen in the $(K, J-K)$ CMD of Baade's 
Window shown in Figure 17 of Frogel \& Whitford (1987). The main stellar sample 
considered by Frogel \& Whitford (1987) consists of M 
giants that were selected from grism surveys, and only two of the stars 
identified as M giants have $J-K > 1.6$. Both of these objects have 
spectral types from Blanco, McCarthy, \& Blanco (1984), 
are photometrically variable, and have 
two $J-K$ colors listed in Table 1 of Frogel \& Whitford (1987). Star \# 87 
(spectral type M8) has $J-K = 1.59$ and 1.76, and so may or may not meet 
the near-infrared color criterion for C 
star status depending on the epoch of observation. Star \# 250 
(spectral type M7) is clearly more interesting, 
as it has $J-K > 2$ in both measurements; however, it is also one of the 
faintest stars classified by Blanco et al.  (1984), and so 
presumably falls near the faint limit of their data.

	The remainder of the stars with $J-K > 1.6$ in the Frogel \& Whitford 
sample are LPVs from the sample that was observed by Glass \& Feast (1982). 
To the best of our knowledge these objects do not have spectra, presumably 
because they are relatively faint at visible wavelengths. In fact, the TCD 
shown in Figure 2 of Frogel \& Whitford (1987) indicates that a star 
with $J-K = 1.6$ will have $V-K \sim 10$. The LPVs in Figure 17 of 
Frogel \& Whitford (1987) tend to have $K_0$ between 6 and 7.5 
(i.e. M$_K$ between --7 and --8.5), and so will have $V_0$ between 
16 and 17.5. After dimming to account for extinction, these stars 
will have $V$ between 17 and 18.5.

	Azzopardi, Lequeux, \& Rebeirot (1985) conducted a grens survey of 
Baade's Window, and found 15 C stars with $V$ between 15.5 and 17.5. 
Azzopardi et al. (1985) argued that these stars are too faint to be C stars 
associated with the Galactic bulge; however, the arguements given 
in the previous paragraph indicate that at least some of the objects 
found by Azzopardi et al. (1985) have $V$ brightnesses that are consistent 
with them being very red C stars at the distance of the Galactic bulge. 
To be sure, some of these objects may be dwarf C stars, which are difficult to 
identify using spectrophotometric information alone (e.g. Downes et al. 
2004). The only ironclad means to confirm the nature of 
the C stars indentified by Azzopardi et al. (1985) is 
to obtain proper motions for these objects, which will allow their 
distances to be measured.

\subsection{The Nature of Intermediate Age Stars Near the Center of M31}

	The detection of intermediate age stars in the ALTAIR data is perhaps 
not a surprise. Spectroscopic studies 
suggest that there is a centrally-concentrated young or intermediate-age 
component in M31 (Bica et al. 1990; Davidge 1997; Sil'chenko et al. 1998). 
The geometric center of M31 has a very red $V-K$ color 
(Davidge et al. 1997), which may be due to a 
population of AGB stars that formed during the past Gyr.

	Studies of the innermost regions of the Galactic 
bulge may provide clues about the nature of the central regions of 
M31. Contrary to expectations (e.g. Figer et al. 2000), 
star formation occurs near the center of the Galaxy. 
There is a cluster of young stars around SgrA* (Lebofsky, Rieke, \& 
Tokunaga 1982; Krabbe et al. 1991; Davidge et al. 1997; 
Blum et al. 2003), as well as other young clusters within the central 100 pc 
(e.g. Cotera et al. 1996; Figer et al. 1999), and at even larger distances 
(Launhardt, Zylka, \& Mezger 2002). The structural properties of the region 
near the Galactic Center (Serabyn \& Morris 1996), the presence of shells at 
large R$_{GC}$ that apparently originate from the inner regions of the Galaxy 
(Bland-Hawthorn \& Cohen 2003), and the analysis of the stellar content near 
the Galactic Center (Blum et al. 2003) together suggest that the current star 
forming activity near the Galactic Center is not a unique event, but 
is likely an on-going or episodic phenomenon with a time scale shorter 
than a Gyr. Nuclei similar to SgrA are also common in nearby spiral galaxies, 
further supporting the arguement that nuclear star formation is episodic 
(e.g. Davidge \& Courteau 2002). 

	The timescale for the disruption of star clusters 
near the Galactic Center is short (e.g. Figer at al. 1999), 
and individual stars may be scattered out of the disk plane by 
interactions with giant molecular clouds (Kim \& Morris 2001). 
Consequently, the stars that form along the plane of the disk 
in the inner regions of a spiral galaxy can 
diffuse away from their natal locations, and be 
injected into the surrounding bulge. Processes of this nature could 
explain the presence of a diffuse intermediate age population 
in the inner Galactic bulge (Wood \& Bessell 1983; 
Harmon \& Gilmore 1988; Lopez-Corredoira, Garzon, \& Hammersley 2001; but see 
also ven der Veen \& Habing 1990). 

	The spatial distribution of the
C stars seen in Bulge 1 and 2 provide clues about their origins. 
If they formed in a nuclear region then it can be anticipated 
that they would be concentrated towards smaller radii. Such a trend 
is not seen in the ALTAIR data, although the uncertainties are high. 
The very red objects in the Stephens et al. (2003) sample, which were 
identified as LPVs in that study, are also uniformly distributed. 
Perhaps the greatest challenge to any model that tries to associate the 
C stars with nuclear star formation comes from the $(K, 
H-K)$ CMDs of the innermost regions of M31 constructed by Davidge (2001). 
These CMDs have relatively tight, well-defined sequences, 
with only a modest number of red stars near the bright end, indicating that 
the number density of C stars may actually drop towards very small 
radii in M31. Given these arguements, it appears that the C stars did 
not form in the nuclear regions of M31, but more likely originated 
over a much larger area.

	Studies of late-type spiral galaxies indicate that the density of 
molecular star-forming material is coupled to disk properties, such as the 
central disk density (Boker, Lisenfeld, \& Schinnerer 2003), rather than 
properties of the inner spheroid, and so it is likely that the C stars in 
Bulge 1 and 2 formed as part of the inner disk of M31, rather than the inner 
spheroid. Nevertheless, the presence of C stars in the inner regions of M31 
suggests that the bulge of this galaxy may be subject to secular processes. 
Kim \& Morris (2001) argue that $1 - 2$ Gyr is required for stars in the inner 
Galaxy to scatter and form an aspect ratio similar to that of the most luminous 
AGB stars in the Galactic Bulge. This is shorter than the upper age limit of 
C stars based on statistical studies (Cole \& 
Weinberg 2002), so it is possible that C star progenitors may survive 
long enough to be redistributed throughout the inner bulge. A radial velocity 
study of the bright, red stars that occur throughout the central few arcmin of 
M31 will provide insight into the nature of these objects.

	We close by noting that with R$_{GC} = 2$ arcmin, Bulge 1 
does not overlap with the central region that has been the target of population 
studies based on integrated spectroscopic information. 
Nevertheless, the spatial distribution of the C stars detected here suggests 
that they are well-mixed throughout the sight line within a few arcmin of 
the center of M31, and so it seems reasonable to expect that some C stars will 
almost certainly occur at smaller R$_{GC}$ than sampled by Bulge 1. 
If this is the case then the C stars seen here may be the brightest members 
of the intermediate-age population that has been detected spectroscopically.

\parindent=0.0cm
\clearpage

\begin{table*}
\begin{center}
\begin{tabular}{lccccccc}
\tableline\tableline
Field Name & RA & Dec & GSC \# & $R_{GC}$ & $\mu_{disk}$ & $\mu_{bulge}$ & Exposures \\
 & E2000 & E2000 & & (arcmin) & & & (seconds) \\
\tableline
Disk 1 & 00:39:13.3 & +40:29:13.3 & 02788-02207 & 61.74 & 22.2 & -- & $20 \times 40$ ($J+H$) \\
 & & & & & & & $24 \times 40$ ($K$) \\
 & & & & & & & \\
Disk 2 & 00:43:21.7 & +41:21:55.2 & 02805-02131 & 9.09 & 20.8 & 20.7 & $24 \times 40$ ($J+H$) \\
 & & & & & & & $27 \times 40$ ($K$) \\
 & & & & & & & \\
Bulge 2 & 00:42:50.9 & +41:12:31.4 & 02801-02008 & 3.83 & 20.5 & 19.2 & $8 \times 40$ ($J+H$) \\
 & & & & & & & $10 \times 40$ ($K$) \\
 & & & & & & & \\
Bulge 1 & 00:42:33.5 & +41:15:50.1 & -- & 2.05 & 20.2 & 18.1 & $8 \times 40$ ($J+H$) \\
 & & & & & & & $10 \times 40$ ($K$) \\
 & & & & & & & \\
\tableline
\end{tabular}
\end{center}
\caption{Field Information and Exposure Times}
\end{table*}

\clearpage

\begin{table*}
\begin{center}
\begin{tabular}{lccccc}
\tableline\tableline
Field name & $\sigma_{HK}^{Measured}$ & $\sigma_{HK}^{Artificial}$ & $\sigma_{JK}^{Measured}$ & $\sigma_{JK}^{Artificial}$ & $\sigma_{JK}^{Residual}$ \\
\tableline
Disk 1 & $\pm 0.14$ & $\pm 0.01$ & $\pm 0.10$ & $\pm 0.04$ & $\pm 0.09$ \\
Disk 2 & $\pm 0.09$ & $\pm 0.02$ & $\pm 0.16$ & $\pm 0.04$ & $\pm 0.15$ \\
Bulge 2 & $\pm 0.11$ & $\pm 0.03$ & $\pm 0.13$ & $\pm 0.07$ & $\pm 0.11$ \\
Bulge 1 & $\pm 0.11$ & $\pm 0.06$ & $\pm 0.22$ & $\pm 0.21$ & $\pm 0.08$ \\
\tableline
\end{tabular}
\end{center}
\caption{Measured and Predicted Dispersions for Stars with $K$ Between 17.5 and 18.5}
\end{table*}

\parindent=0.0cm

\clearpage

\begin{center}
FIGURE CAPTIONS
\end{center}

\figcaption
[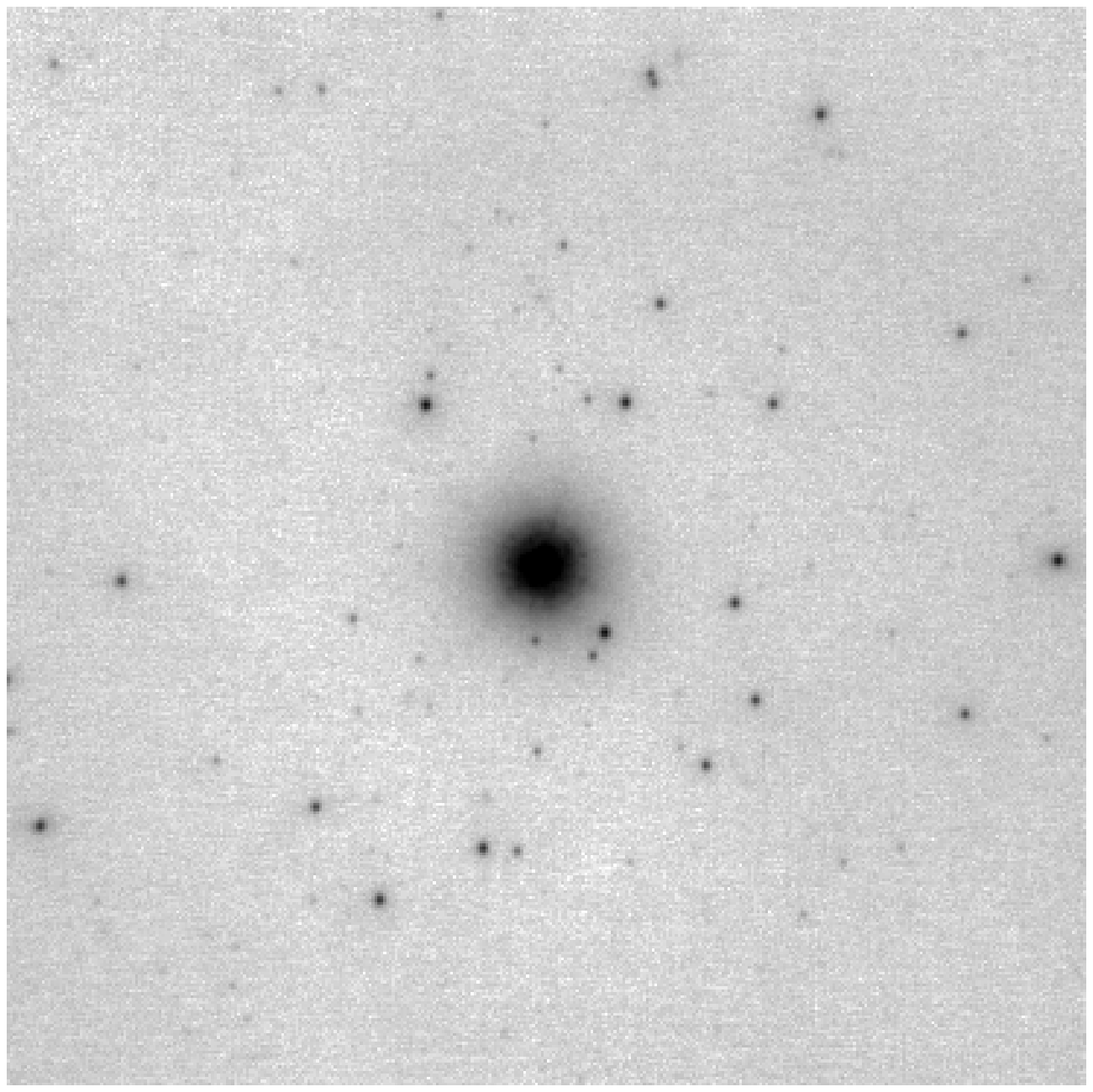]
{The final $K'$ image of the Disk 1 field. The displayed 
field is roughly $17.5 \times 17.5$ arcsec.}

\figcaption
[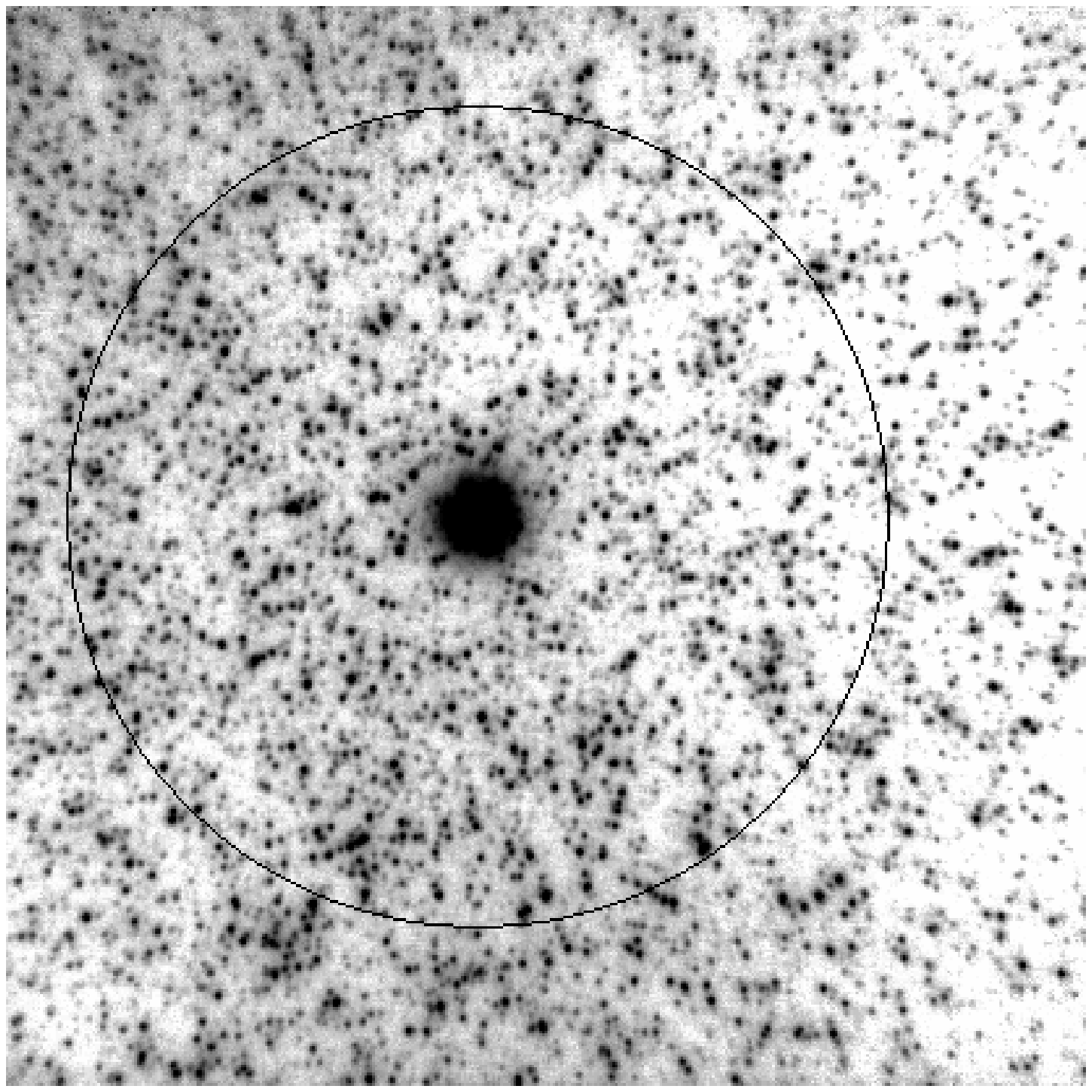]
{The final $K'$ image of the Bulge 1 field. The displayed 
field is roughly $17.5 \times 17.5$ arcsec. The circle shows the outer 
boundary of the region used in the photometric analysis.}

\figcaption
[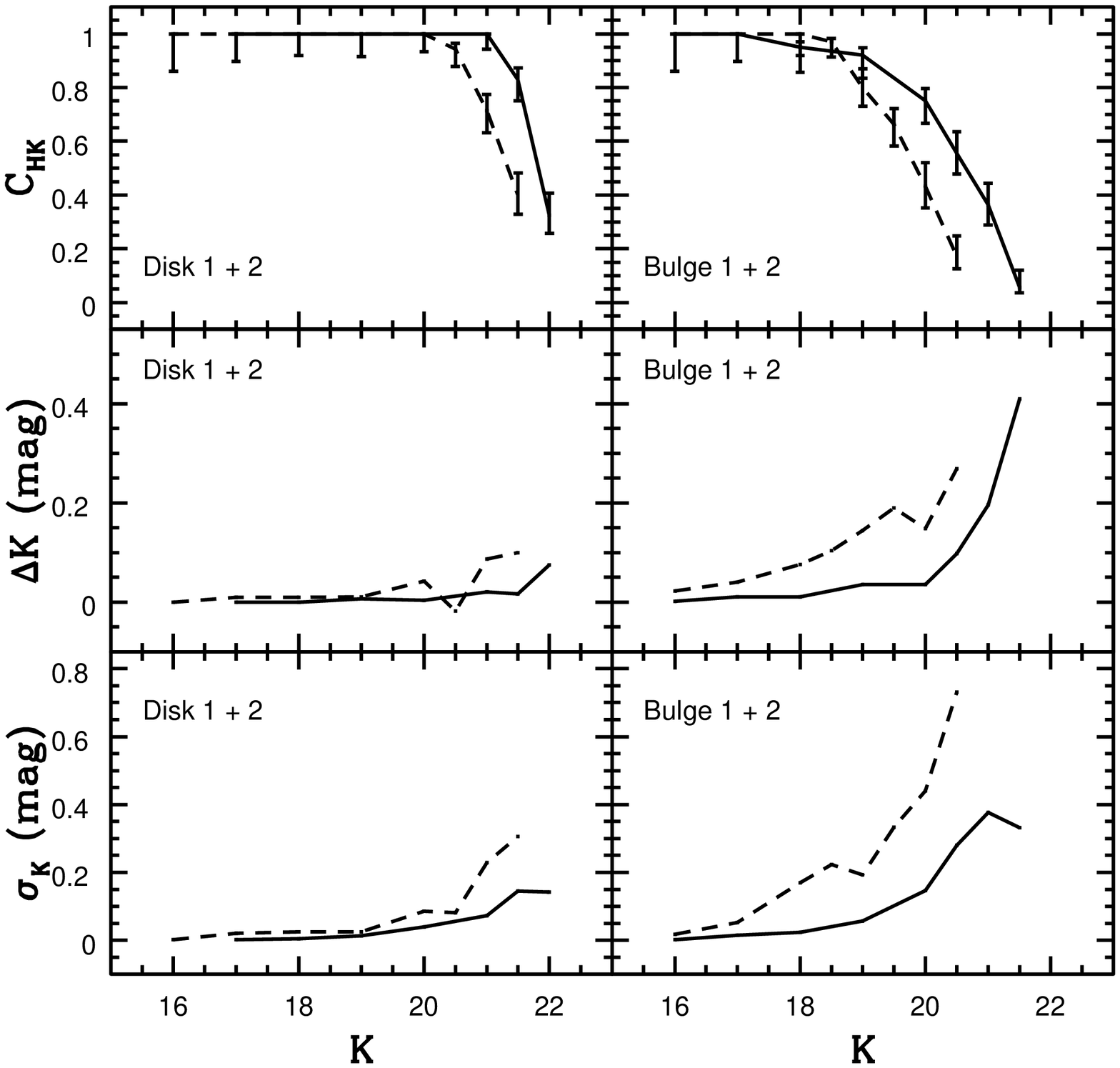]
{The results of artificial star experiments in the $K-$band. Shown are (1) the 
completeness fraction of stars detected in both $H$ and $K$, $C_{HK}$, (2) the 
mean difference between the actual and measured brightnesses, $\Delta$K, 
and (3) the estimated uncertainty in the photometric measurements, $\sigma_K$. 
$C_{HK}$ is the the number of artificial stars recovered divided by the 
number that were added to the frame in both filters, while $\sigma_K$ is the 
standard deviation about $\Delta$K. The panels in the left hand column 
show the results for Disk 1 (solid line) and 
2 (dashed line), while the panels in the right hand column show the results 
for Bulge 1 (dashed line) and 2 (solid line). The error bars show the 
$1-\sigma$ uncertainties in the completeness measurements.}

\figcaption
[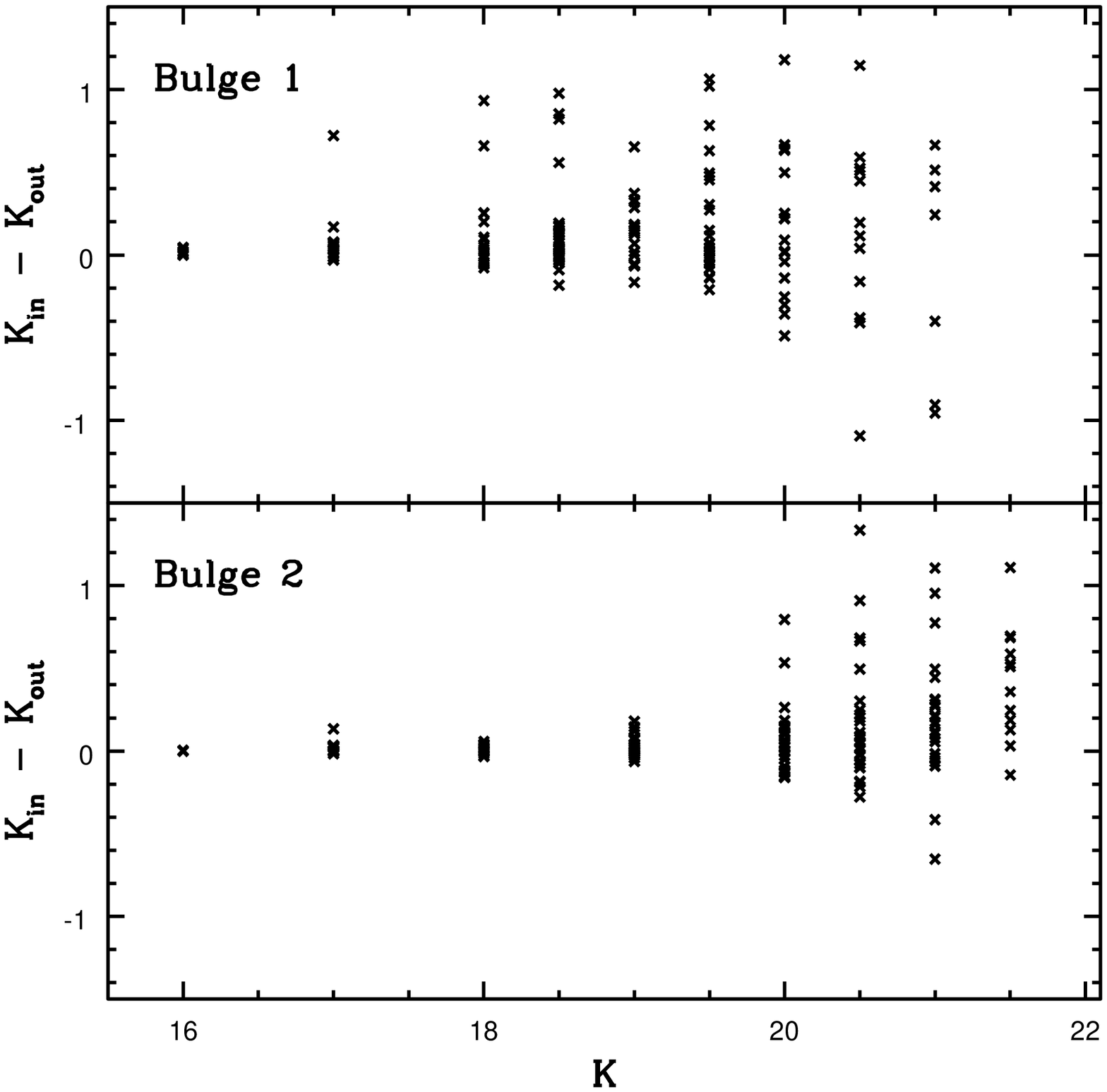]
{The difference between the actual and measured brightnesses of artificial 
stars, K$_{in}$ -- K$_{out}$, for Bulge 1 and 2. Note that blending only 
becomes significant in Bulge 2 when $K = 20$. As for Bulge 1, blending becomes 
significant when $K > 19$, although the simulations indicate that 
a modest fraction of stars with $K > 18$ may be blends.}

\figcaption
[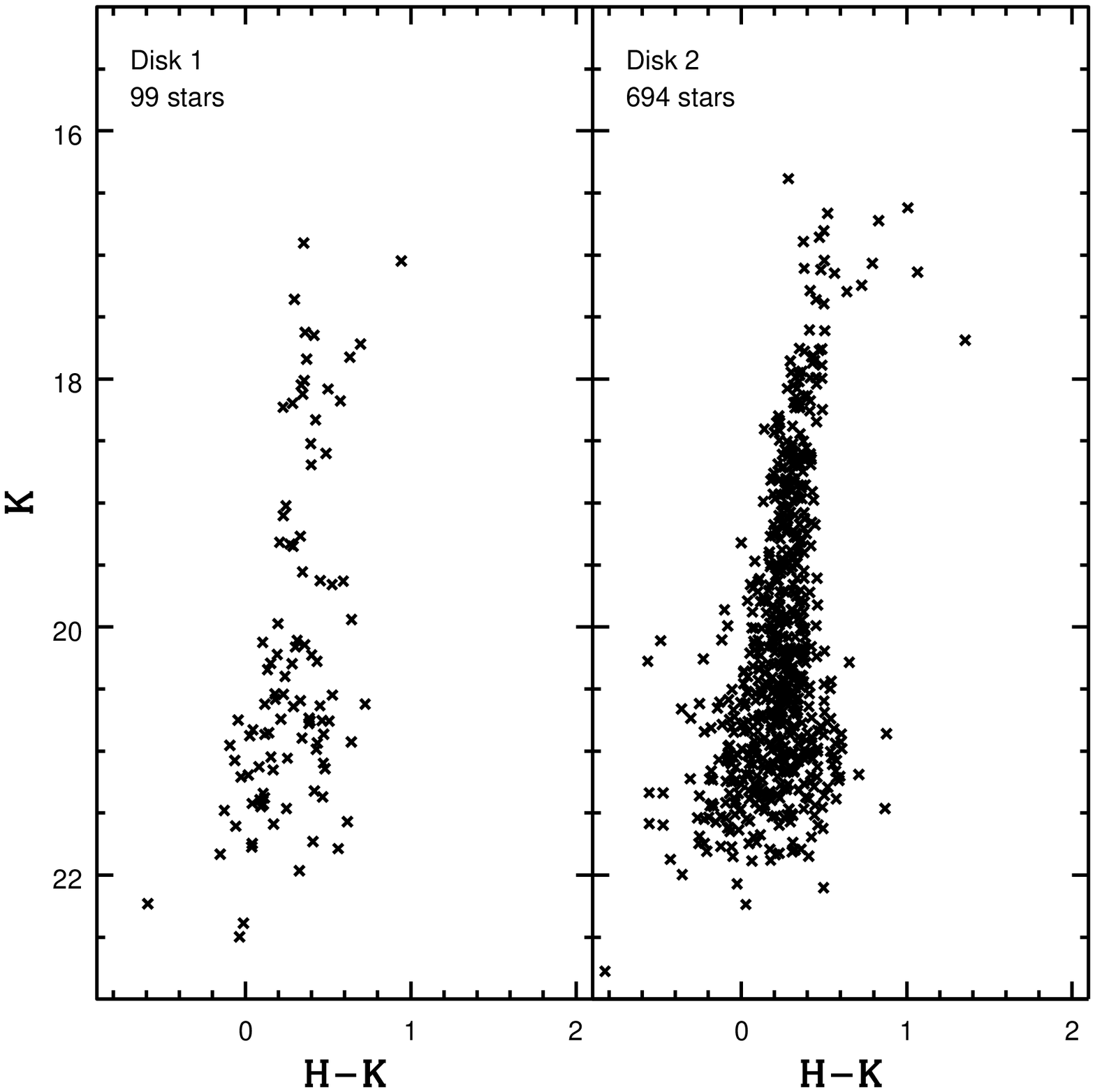]
{The $(K, H-K)$ CMDs of Disk 1 and 2. The number 
of points plotted in each CMD is given in the upper left hand corner of 
each panel.}

\figcaption
[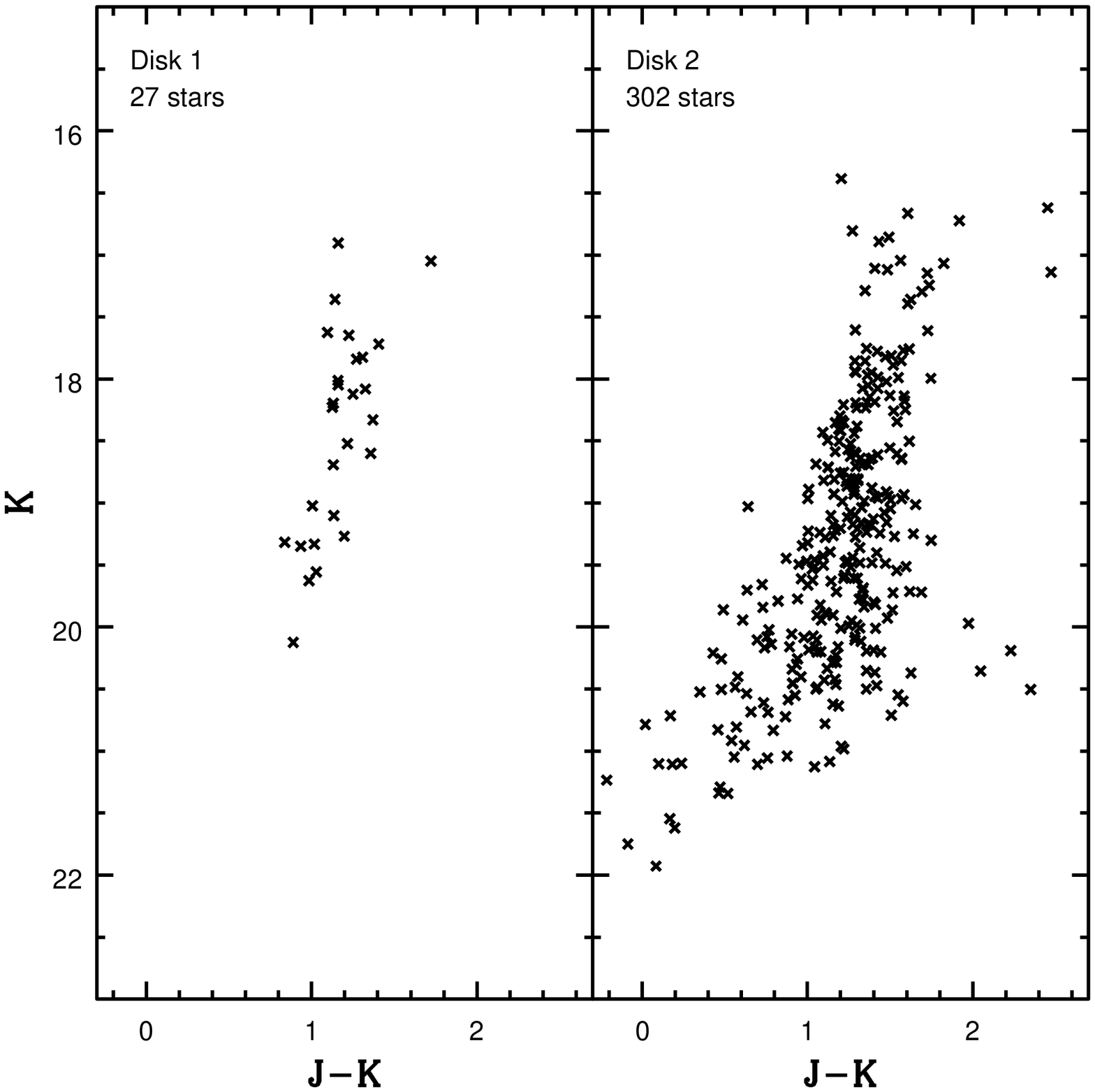]
{The $(K, J-K)$ CMDs of Disk 1 and 2. The number 
of points plotted in each CMD is given in the upper left hand corner of 
each panel.}

\figcaption
[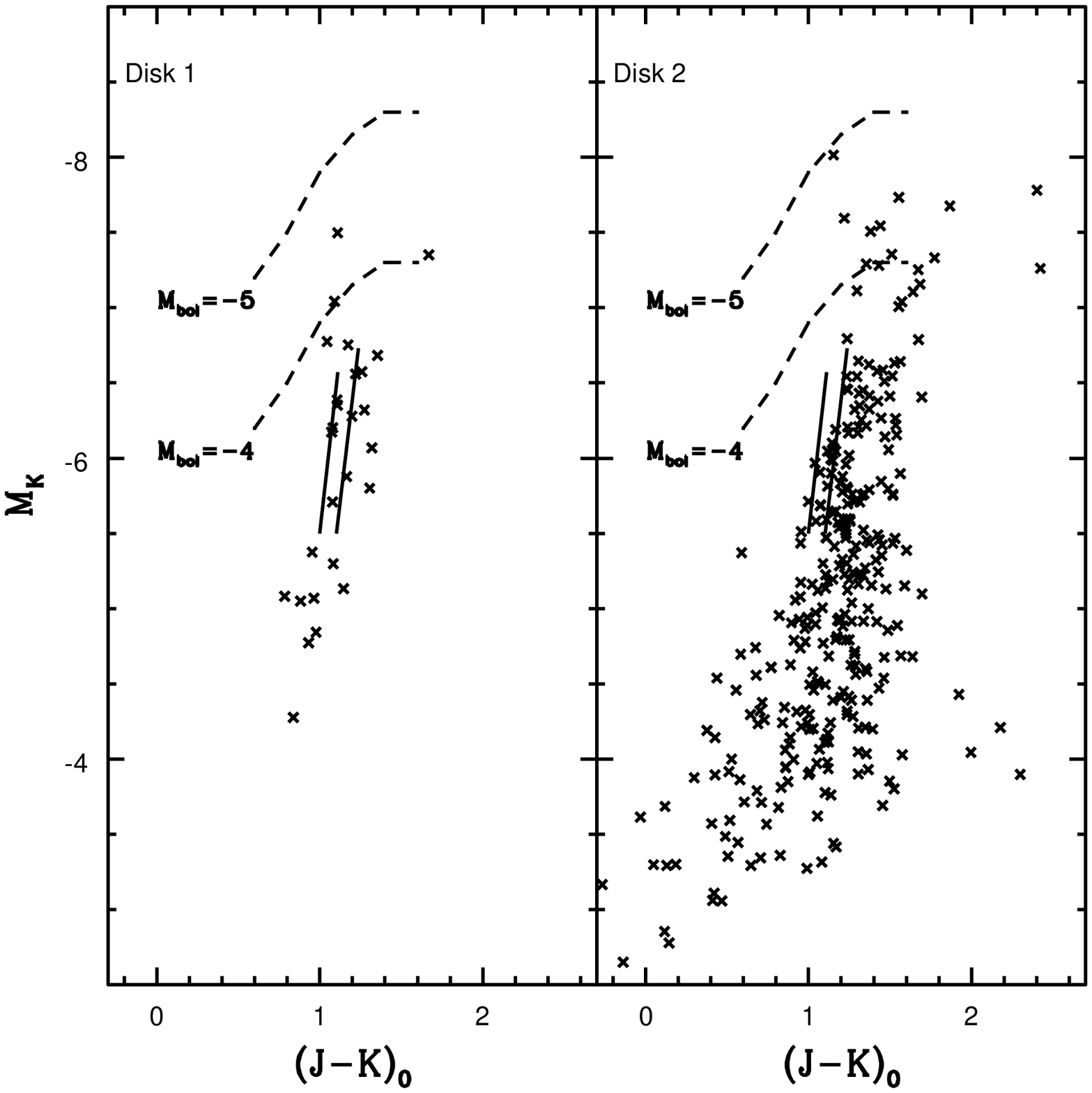]
{The $(M_K, (J-K)_0)$ CMDs of Disk 1 and 2. A 
distance modulus of 24.4 and a foreground reddening of E(B--V) = 0.1 have been 
assumed. The solid lines are the RGB sequences for the globular clusters
47 Tuc and NGC 6528, based on the fiducial colors listed in Table 2 of Ferraro 
et al. (2000). The dashed lines show sequences for M$_{bol} = -4$ and --5, 
which were calculated using the bolometric corrections for Galactic bulge 
giants shown in Figure 1b of Frogel \& Whitford (1987).}

\figcaption
[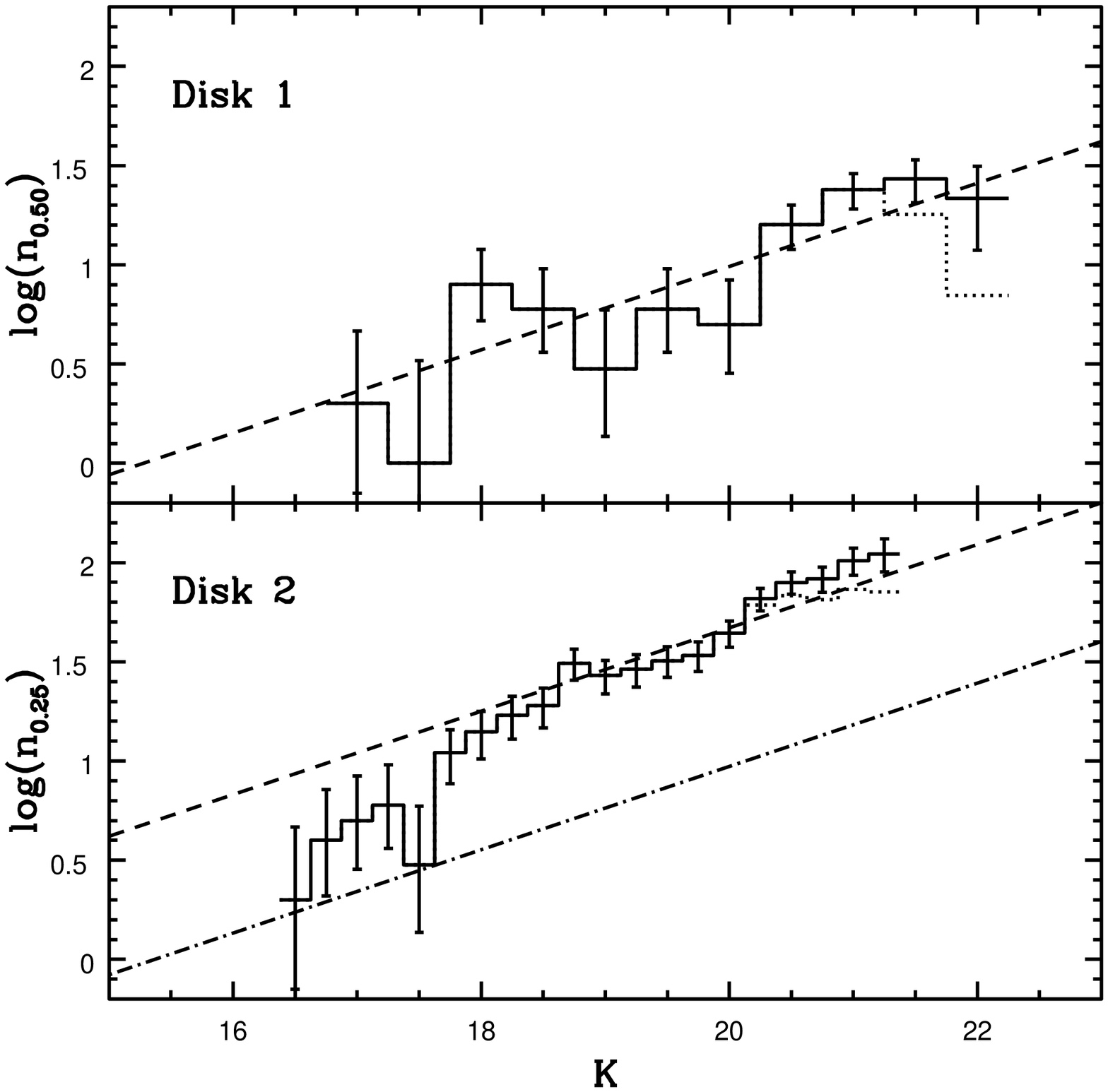]
{The $K$ luminosity functions of Disk 1 and 2. 
The solid line shows the LF corrected for incompleteness, while the 
dotted line shows the raw LF. $n_{0.5}$ and $n_{0.25}$ are the number of stars 
per 0.5 and 0.25 magnitude intervals in $K$; the relatively coarse 0.5 
mag interval was used in Disk 1 to reduce the noise per bin. 
The errorbars show the $1-\sigma$ uncertainties due to Poisson statistics, 
as computed by Gehrels (1986), added in quadrature to the uncertainties 
in the completeness corrections, computed using binomial statistics. 
The dashed line shows a power-law with exponent 0.21, computed from the 
RGB LF of the globular cluster NGC 6528, that was 
fit to the LF bins with $K$ between 18 and 21; the 
dotted-dashed line in the bottom panel shows the relation fit to the 
Disk 2 data but shifted down by log(0.2) $ = -0.7$ units to approximate the 
number counts expected for a pure AGB population. Note that the discontinuity 
due to the RGB-tip can clearly be seen in Disk 2 near $K = 17.75$.}

\figcaption
[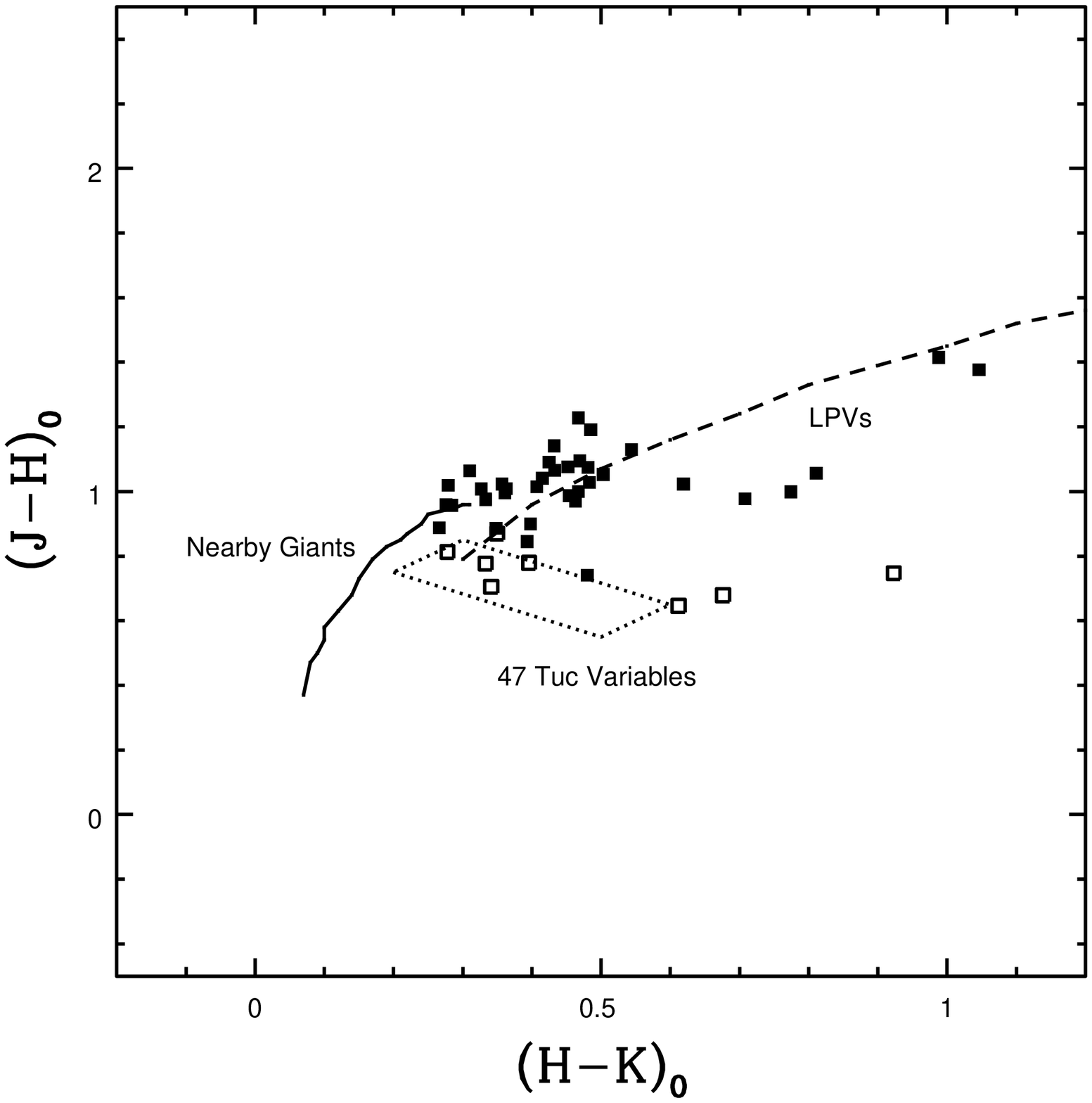]
{The $(J-H, H-K)$ two-color diagram of stars in Disk 1 (open 
squares) and 2 (filled squares). Only stars with $K < 18$ 
are plotted to reduce scatter and clutter, and colors have 
been corrected only for foreground reddening. The solid line shows the locus 
of solar neighborhood giants from Bessell \& Brett (1988), while the dashed 
line shows the locus of LPVs in the SMC and LMC based on data from Wood et 
al. (1983, 1985). The dotted line shows the region in Figure 7 of 
Frogel, Persson, \& Cohen (1981) that is occupied by the 
variable stars V1, V2, and V3 in 47 Tuc.}

\figcaption
[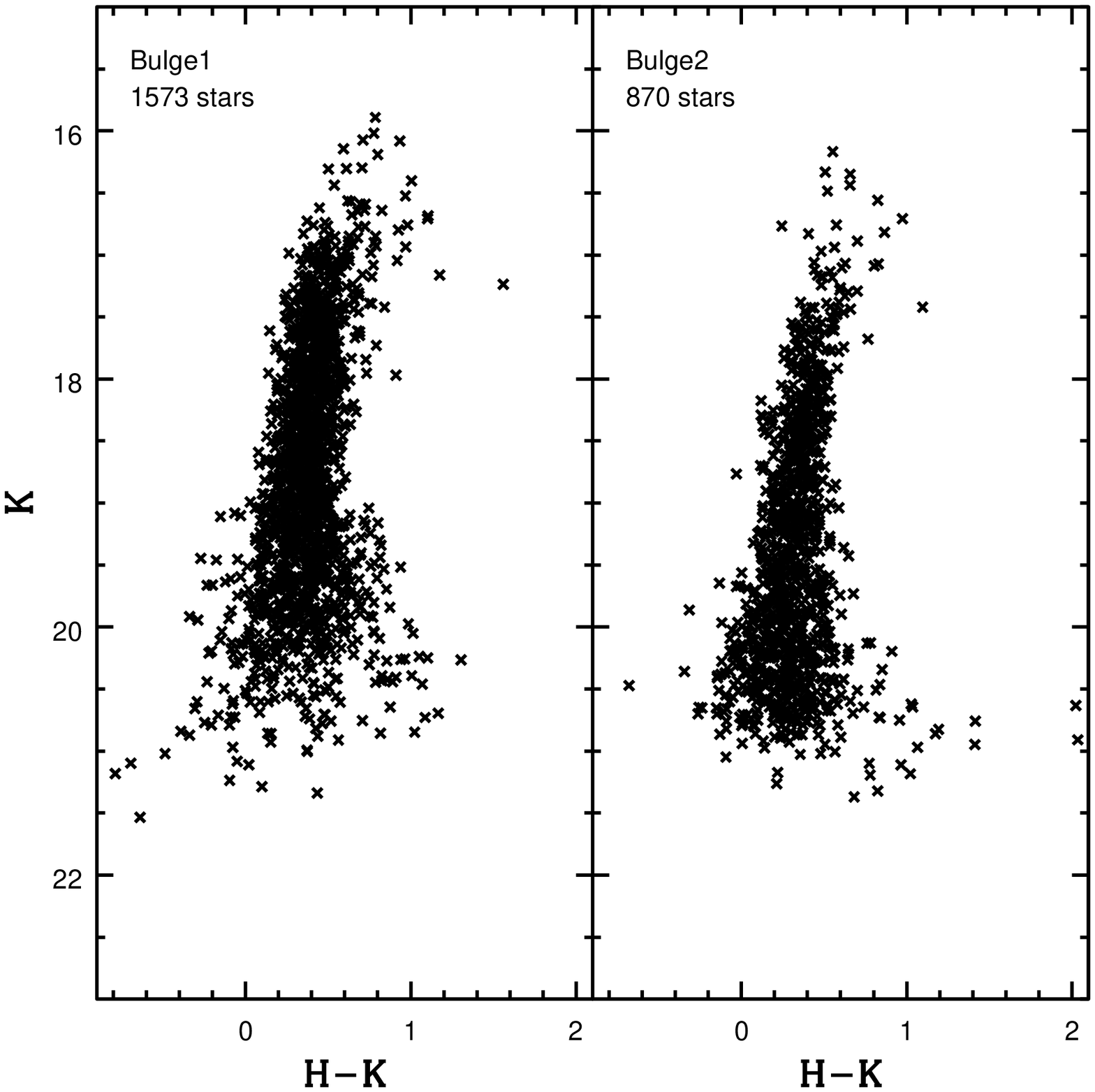]
{The $(K, H-K)$ CMDs of the bulge fields. The number 
of points plotted in each CMD is given in the upper left hand corner of 
each panel.}

\figcaption
[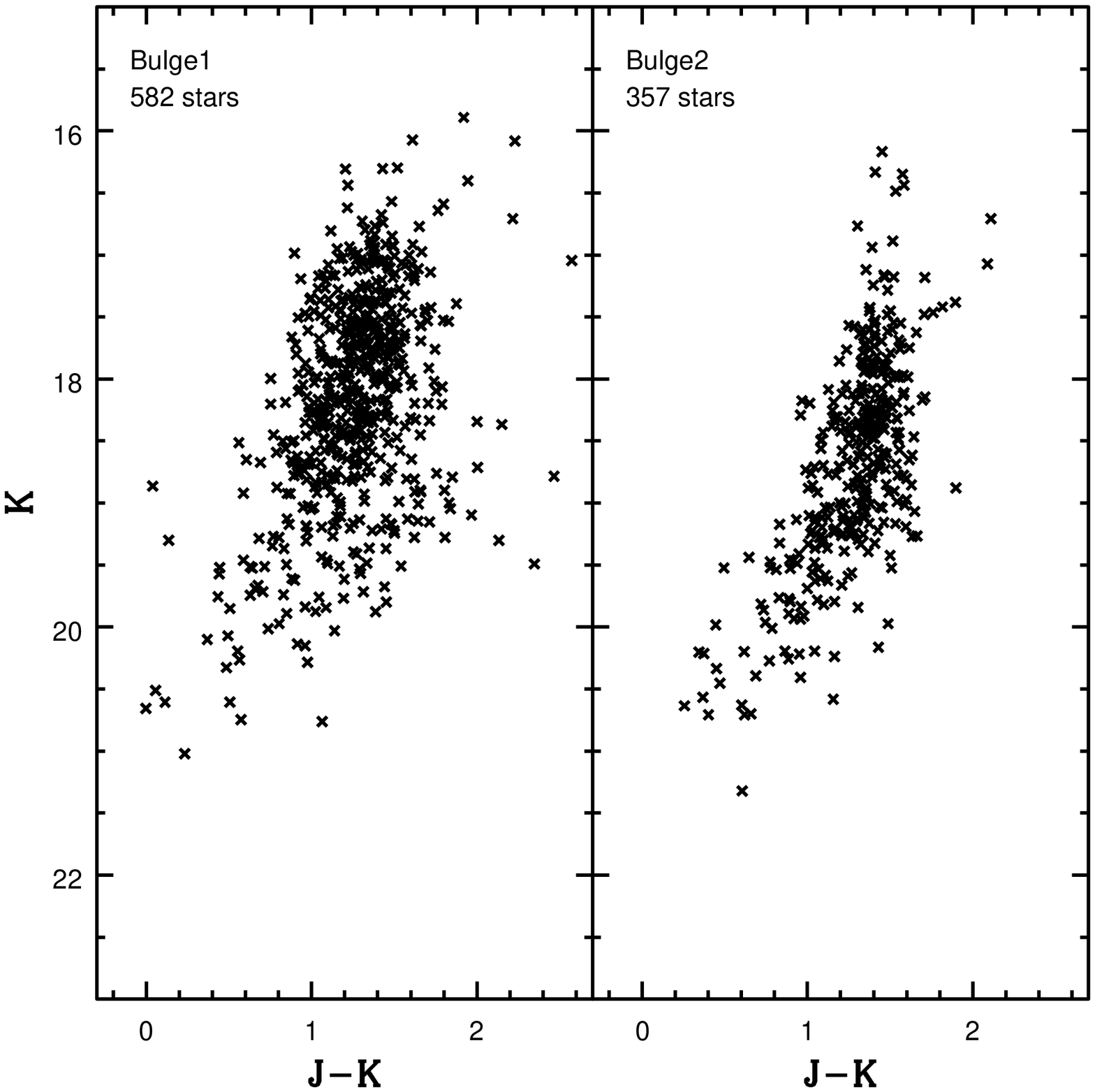]
{The $(K, J-K)$ CMDs of the bulge fields. The number 
of points plotted in each CMD is given in the upper left hand corner of 
each panel.}

\figcaption
[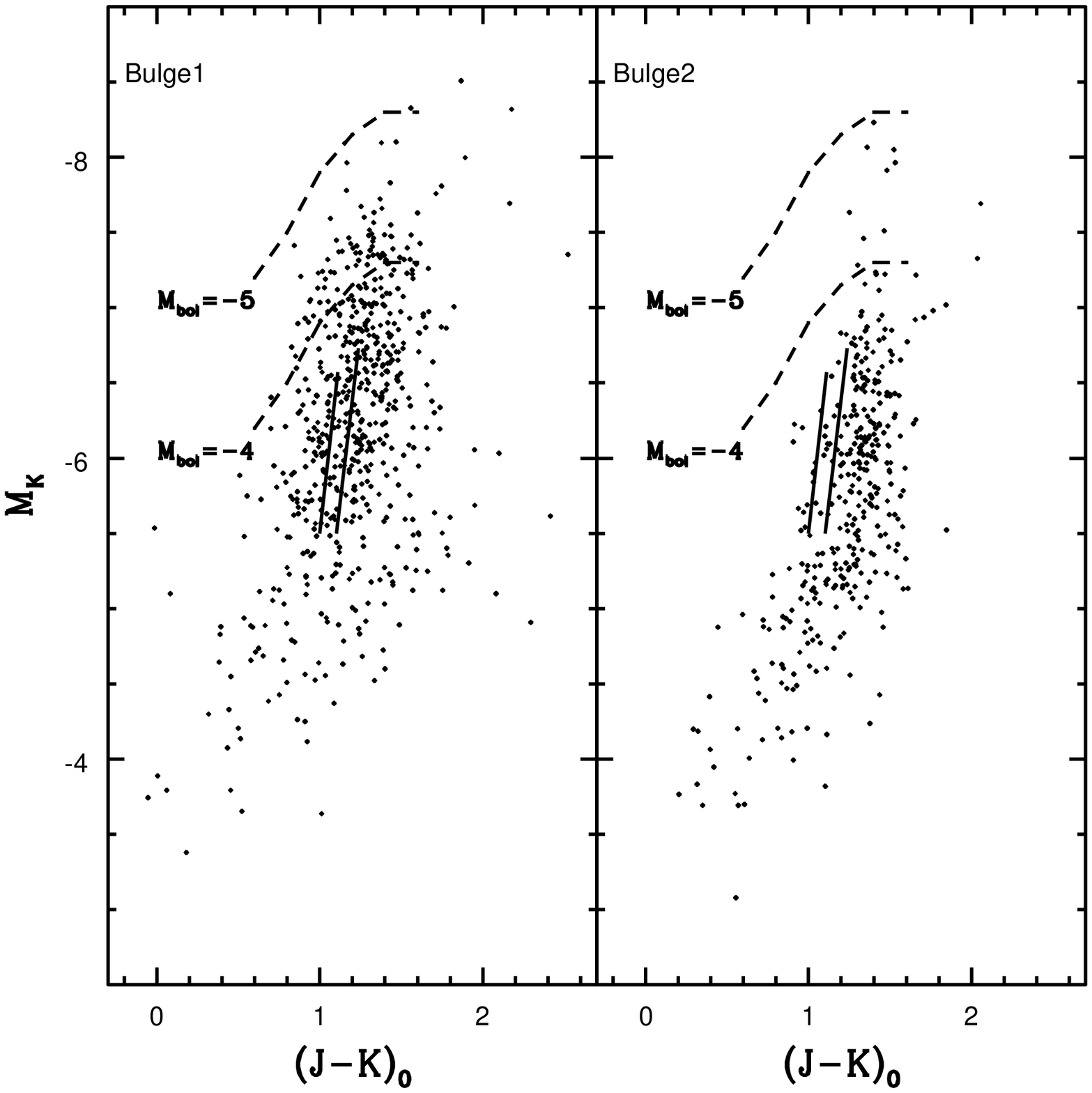]
{The $(M_K, (J-K)_0)$ CMDs of the bulge fields. A 
distance modulus of 24.4 and a foreground reddening of E(B--V) = 0.1 have been 
assumed. The solid lines show the RGB sequences for the globular clusters
47 Tuc and NGC 6528, based on the fiducial colors listed in Table 2 of Ferraro 
et al. (2000). The dashed lines show sequences for M$_{bol} = -4$ and --5 
that were computed using the bolometric corrections for Galactic bulge giants 
shown in Figure 1b of Frogel \& Whitford (1987).}

\figcaption
[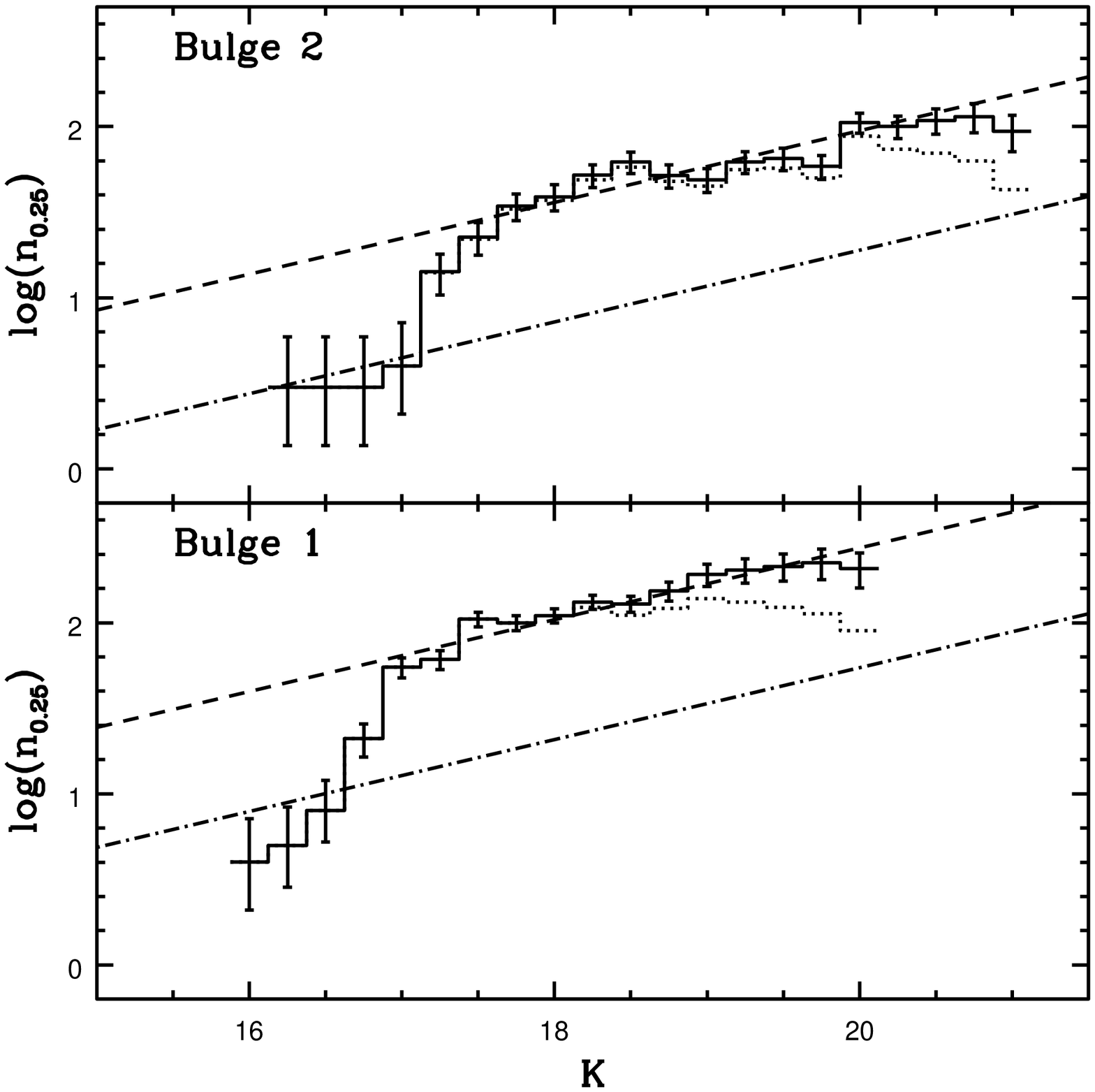]
{The $K$ luminosity functions of the bulge fields.
The solid line shows the LF corrected for incompleteness, while the 
dotted line shows the raw number counts. $n_{0.25}$ is the number of stars in 
the field per 0.25 magnitude interval in $K$. 
The errorbars show the $1-\sigma$ uncertainties due to Poisson statistics, 
as computed by Gehrels (1986), added in quadrature to the uncertainties 
in the completeness corrections, computed using binomial statistics. 
The dashed line shows a power-law with exponent 0.21, computed from the 
LF of RGB stars in the globular cluster NGC 6528, that has been fitted 
to the LF entries with $K$ between 18 and 20. The dashed-dotted line 
shows this same power-law relation, but with the y-intercept decreased by 80\%, 
to approximate the trend expected for stars evolving on the AGB. Note that (1) 
the RGB-tip in Bulge 2 appears to be 0.25 -- 0.50 mag fainter than in 
Bulge 1, and (2) the numbers of very bright stars in both fields are 
consistent with evolution on the AGB.}

\figcaption
[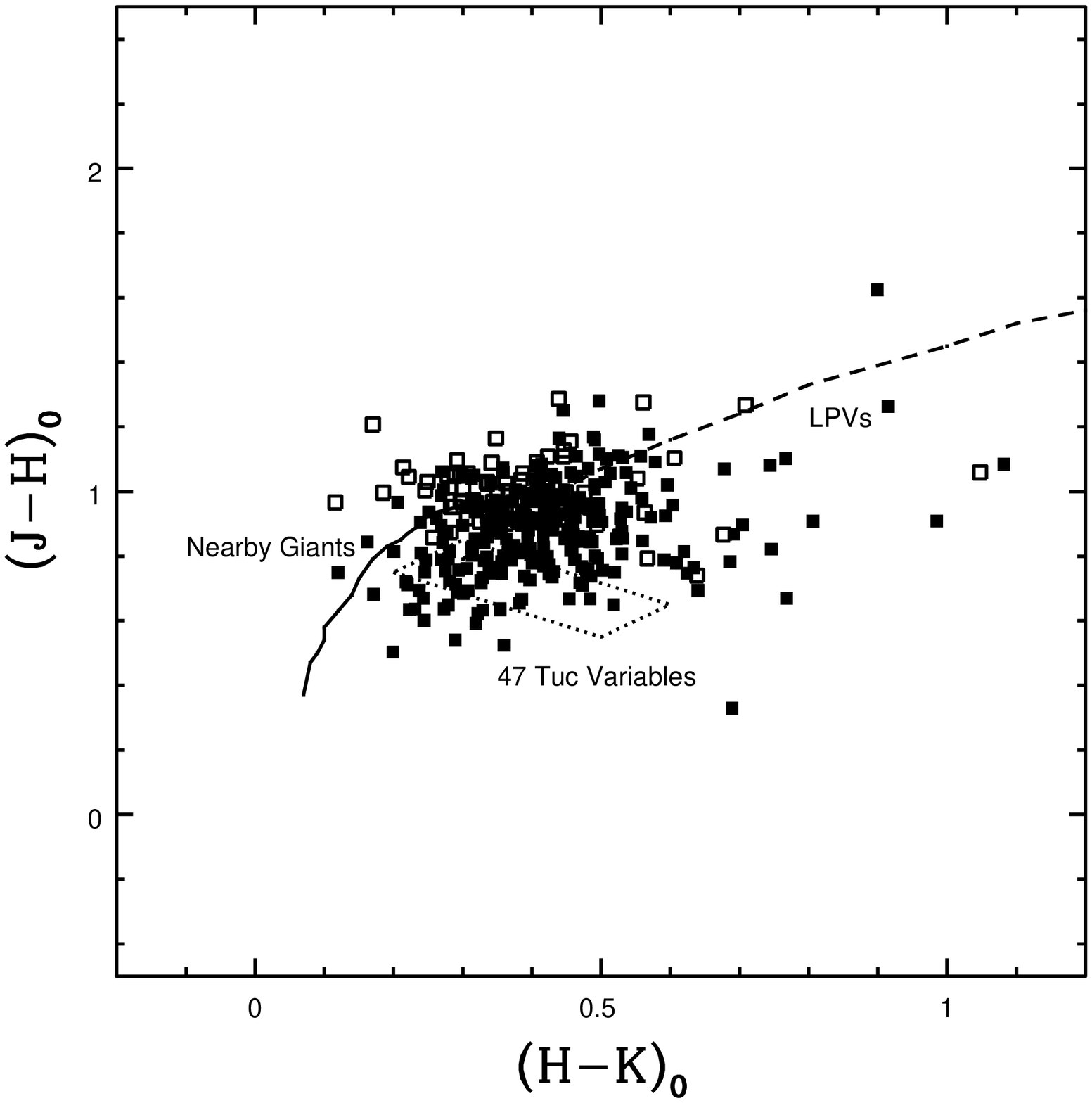]
{The $(J-H, H-K)$ two-color diagram of stars in Bulge 1 (open 
squares) and 2 (filled squares). Only stars with $K < 18$ 
are plotted to reduce scatter and clutter, and the colors have 
been corrected only for foreground reddening. The solid line shows the locus of 
solar neighborhood giants from Bessell \& Brett (1988), while the dashed 
line shows the locus of LPVs in the SMC and LMC based on data from Wood et 
al. (1983, 1985). The dotted line shows the portion of the TCD occupied by the 
bright variable stars V1, V2, and V3 in 47 Tuc, based on Figure 7 of Frogel, 
Persson, \& Cohen (1981).}

\figcaption
[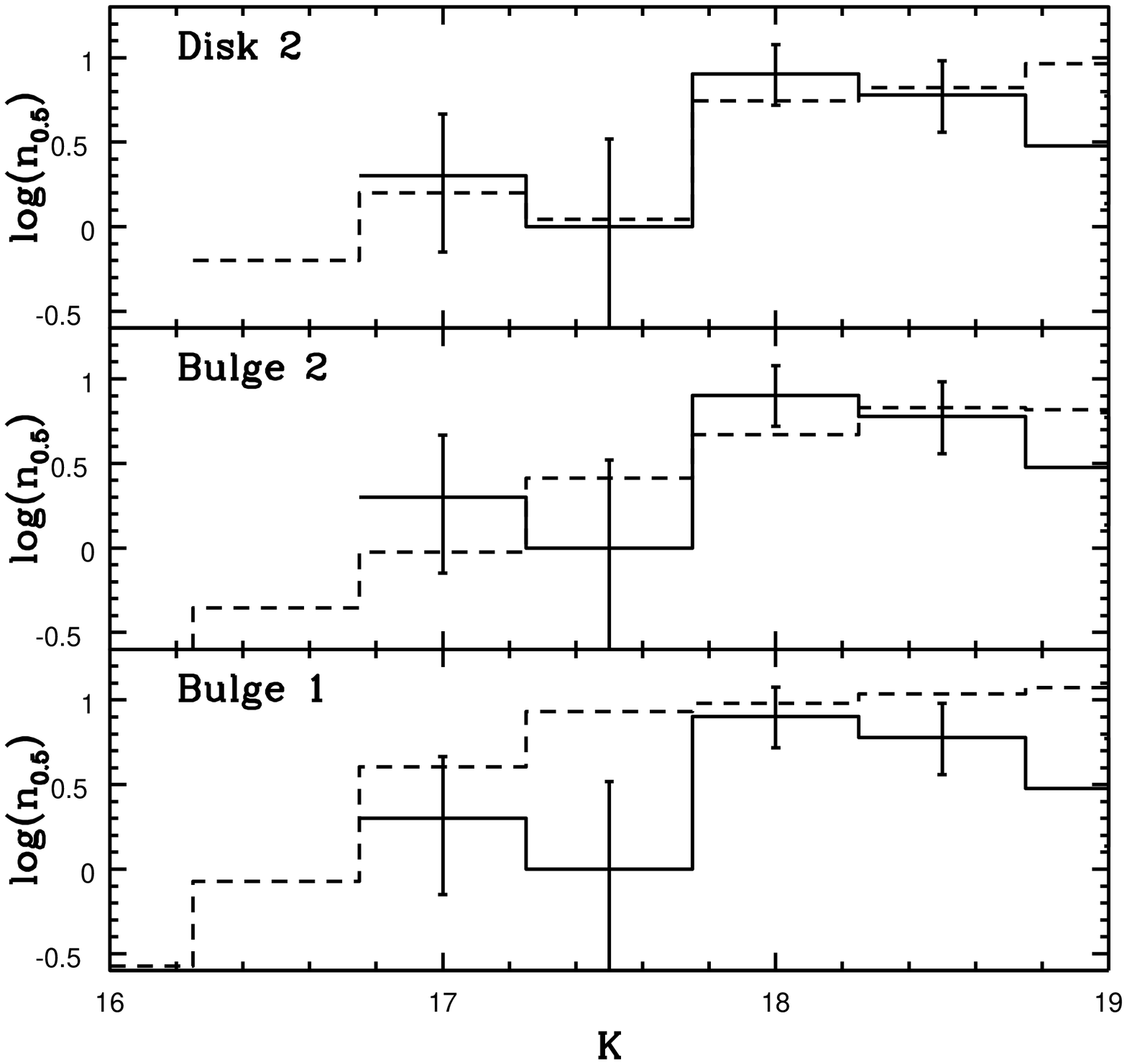]
{The $K$ LFs of the Disk 2, Bulge 2, and Bulge 1 fields 
(dashed lines), compared with the $K$ LF of Disk 1 
(solid line). The Disk 2, Bulge 1, and Bulge 2 LFs 
have been scaled to match the density of stars in Disk 1 
using $r-$band surface brightness measurements from Kent (1987). 
n$_{0.5}$ is the number of stars per 0.5 magnitude interval in Disk 1. 
The errorbars show the $1-\sigma$ uncertainties due to Poisson statistics, 
as computed by Gehrels (1986), added in quadrature to the uncertainties 
in the completeness corrections, computed using binomial statistics. 
Note that the Disk 1 LF matches that of the Disk 2 
and Bulge 2 at the $1-\sigma$ level over a wide range of brightnesses, 
suggesting that the number of blends in the Disk 2 and Bulge 2 fields 
are small. While the Bulge 1 LF typically also agrees with the Disk 1 
LF within the $1-\sigma$ level, the Bulge 1 measurements fall 
systematically above those in Disk 1.}

\figcaption
[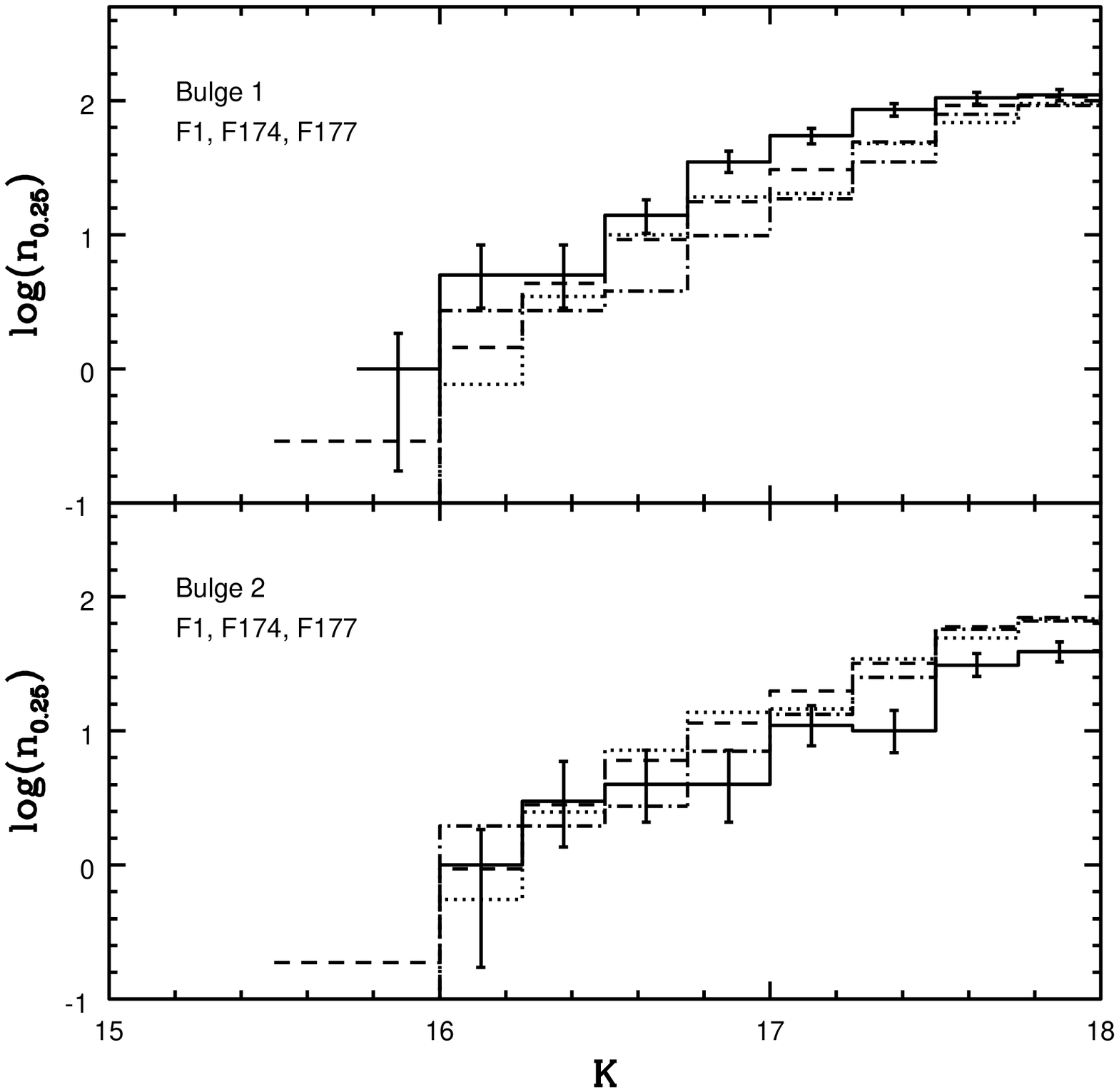]
{The $K$ LF of Bulge 1 and 2 (solid lines) compared with the LFs of Stephen et 
al. (2003) F1 (dashed line), F174 (dash-dot line), and F177 (dotted line). 
$n_{0.25}$ are the number of stars per 0.25 mag interval in Bulge 1 and 2. 
The errorbars show the $1-\sigma$ uncertainties due to Poisson statistics, 
as computed by Gehrels (1986), added in quadrature to the uncertainties 
in the completeness corrections, computed using binomial statistics. 
The Stephens et al. (2003) data have been scaled to account for differences in 
(1) surface brightness, using the $r-$band measurements from Kent (1987), and 
(2) area covered.}

\end{document}